\documentclass{aa}

\usepackage{graphicx}
\usepackage{txfonts}
\usepackage[draft]{hyperref}

\newcommand{\pristine}{Martin, Starkenburg et al.}

\begin{document} 

  \renewcommand\sectionautorefname{Sect.}
  \renewcommand\subsectionautorefname{Sect.}
  \renewcommand\subsubsectionautorefname{Sect.}
  \renewcommand\figureautorefname{Fig.}

  \title{The Pristine Survey} 
  \subtitle{XXX. Mapping substructures in the inner and outer halo
            with blue horizontal branch stars}
  \author{Manuel~Bayer\inst{1}  \and Else Starkenburg\inst{1}
          \and Annemarijn~Zwerver\inst{1, 2}
          \and  Fraser~McKibben\inst{1} \and Maya Bohak\inst{1}
          \and Nicolas~F.~Martin\inst{3, 4}  \and Emma~Dodd\inst{5}
          \and Hanneke~C.~Woudenberg\inst{1}
          \and Guillaume~F.~Thomas\inst{6, 7}
          \and Amina~Helmi\inst{1} \and Julio~F.~Navarro\inst{8, 9}}
  \institute{Kapteyn Astronomical Institute, University of
             Groningen, Landleven 12, NL-9747AD Groningen, the
             Netherlands\\
             \email{mbayer@astro.rug.nl}
             \and Utrecht University, Faculty of Geosciences,
             Utrecht, the Netherlands
             \and Universit{\'e} de Strasbourg, Observatoire
             astronomique de Strasbourg, UMR 7550, F-67000
             Strasbourg, France
             \and Max-Planck-Institut f{\"u}r Astronomie, 
             K{\"o}nigstuhl 17, D-69117 Heidelberg, Germany
             \and Institute for Computational Cosmology, Department
             of Physics, Durham University, South Road, Durham, DH1
             3LE, UK
             \and Instituto de Astrof{\'i}sica de Canarias, E-38205 
             La Laguna, Tenerife, Spain
             \and Universidad de La Laguna, Dpto. Astrof{\'i}sica, 
             E-38206 La Laguna, Tenerife, Spain
             \and Department of Physics and Astronomy, University of
             Victoria, Victoria, BC, Canada
             \and KU-KIST Green School, Korea University}
  \abstract%
  {Blue horizontal branch (BHB) stars serve as good standard candles
   and kinematic tracers of the Milky Way halo. The halo displays
   various overdensities and kinematical substructures of different
   shapes and sizes.}%
  {Our objective is to investigate the substructure in three
   dimensions and the bulk motion of the halo up to distances well
   exceeding the limit of the \textit{Gaia} parallaxes.}
  {We selected BHB stars in the $\left(u-CaHK\right) _ 0$-$(g-r) _ 0$
   colour-colour diagram with a hierarchical Bayesian
   mixture model of three populations: BHB stars, blue straggler
   stars, and outliers.}%
  {Using $CaHK$ photometry derived from \textit{Gaia} XP spectra and
   the Pristine survey, we provide both a shallow and a deep
   catalogue of BHB stars of 7,057 and 8,824 stars, respectively,
   where the deep catalogue is restricted to the deeper Pristine
   footprint compared to the shallower with \textit{Gaia} XP. Using
   the shallow catalogue, we trace to larger distances 6D
   substructures identified earlier within a 2.5 kpc from the Sun,
   including accepted populations; namely, Gaia-Enceladus, Thamnos,
   Helmi streams, Sequoia, and other smaller halo substructures such
   as ED-3, ED-4, and ED-6. The deep catalogue provides reliable
   distance measurements for a more detailed 3D mapping of the
   Sagittarius stream and of the Pisces Overdensity. We add new
   candidate BHB stars with line-of-sight velocities to existing
   catalogues and confirm previously reported differences between the
   northern and southern Galactic hemispheres, likely due to the infall
   of the Large Magellanic Cloud.}%
  {This research demonstrates that probabilistic photometric
   classification can effectively identify distant candidate BHB
   stars, and can provide catalogues of these stars that enhance our
   understanding of the 3D and 6D structures in the halo of the
   Milky Way.}

  \titlerunning{Pristine Survey blue horizontal branch stars}
  \authorrunning{Manuel~Bayer et al.}
  \maketitle
  \nolinenumbers

  \section{Introduction}\label{intro}

  The halo comprises only $\sim 1$ percent of the Milky Way's total
  stellar mass
  \citep[e.g.][and references therein]{BlandHawthorn2016} and yet
  it has received increased attention over the past three
  decades, mainly because it preserves an archaeological record of
  the Milky Way's accretion and merging history
  \cite[e.g.][]{helmi_and_simon_white1999, paul_harding_et_al2001,
                massari_et_al2019, Dodd_2023}.
  Even though most of the accreted mass has been phase-mixed
  \cite[e.g.][]{de_lucia_and_helmi2008, helmi_et_al2018,
                horta_et_al2023astrophysical_journal,
                genina_et_al2023, alex_h_riley_et_al2025},
  much of its structure is preserved in the space of dynamical
  quantities (e.g.
  \citealt{helmi_and_de_zeeuw2000, helmi_et_al2018,
           koppelman_et_al2019_631_l9, Dodd_2023}).
  Smaller, and/or more recent accretion or disruption events are
  additionally still visible on the sky as narrow stellar streams
  \citep[e.g.][]{Ibata2024}. Chemical abundances can significantly
  help to identify progenitors and their properties
  \cite[e.\,g.][]{helmi_et_al2018, fernandez_alvar_et_al2018_apj,
                  Dodd2025}. 

  Thanks to the \textit{Gaia} mission \citep{gaiarelease} and to the
  completion of large spectroscopic surveys, numerous overdensities
  and kinematical substructures have been found in the halo
  \cite[e.g.][]{helmi_et_al2018, koppelman_et_al2019_631_l9,
                Naidu_2020, L_vdal_2022, oria_et_al2022antaeus,
                Dodd_2023}.
  In fact, most of the mass in the stellar halo originates from a few
  massive merging events
  \cite[e.g.][]{pillepich_et_al2018_475_648,
                monachesi_et_al2019, Deason_2019, Helmi_2020,
                Naidu_2020, orkney_et_al2022_517, genina_et_al2023,
                guillaume_f_thomas_et_al2025},
  including Gaia-Enceladus
  ($\text{stellar mass}\sim 10^8$--$10^9 M_{\odot}$,
  \citealt{belokurov_et_al2018gaia_sausage, helmi_et_al2018, 
           fernandez_alvar_et_al2018_apj,
           mackereth_et_al2019_482_3426m}).

  While 6D position and velocity information is available for stars in
  the solar neighbourhood, standard candles are indispensable for the
  measurement of the 3D structure at larger distances in the Milky Way
  stellar halo, in particular for the outermost regions. Parallax
  measurements decrease in accuracy beyond the solar neighbourhood
  \citep{gaiadr3summary}, and therefore thorough 3D (or
  6D phase space when velocities are available) characterisation is
  most reliably restricted within the order of a ten-kiloparsec radius
  \citep[e.g.][]{Dodd_2023}. On the other hand, standard candles,
  including RR Lyrae, M-giants, carbon stars, and blue horizontal
  branch (BHB) stars are reliable distance indicators that can be used
  to trace larger distances
  \cite[e.g.][]{Yanny_2000, majewski_et_al2003, sesar_et_al2017,
                Deason2011, deason_et_al2012, muraveva_et_al2025}.

  In this study, our objective is to select BHB stars to study
  substructure in the Galactic halo. BHB stars are a component of the
  halo's ancient stellar population. They are defined as hot and bright
  A-type stars that have advanced further along their stellar
  evolution, are helium-burning, and possess lower masses compared to
  main-sequence A-type stars \citep{Yanny_2000}. Their absolute
  magnitudes of $M_{g} \sim 0.5$--0.7
  \citep{Yanny_2000, Deason2011, barbosa_et_al2022} in the Sloan
  Digital Sky Survey \citep[SDSS,][]{sdss} \textit{g} band render
  these stars discernible out to considerable distances, well
  beyond the reach of current parallax measurements and out into the
  outer halo ($R_{GC} > 20 $ kpc). In some studies, BHB stars have been
  used to provide 3D information on (relatively unexplored) regions of
  the Galaxy beyond 100\,kpc (e.g.
  \citealt{Thomas2018Atype, tetsuya_fukushima_et_al2019,
           Starkenburg2019, amarante_et_al2024}).

  Because of their colours, photometric searches of BHB stars
  contain few contaminants, mainly blue straggler stars. Hot
  ($6500 \lesssim T_\text{eff}\,\text{K}^{-1} \lesssim 20000$) and young
  main-sequence stars are very rare in the stellar halo because of
  its older stellar populations \citep{mihalas_and_binney1981},
  while other hot objects such as quasars, white dwarfs, and hot
  subdwarfs can be eliminated by careful broad-band colour cuts
  \cite[e.g.][]{lenz_et_al1998, Yanny_2000, Deason2011,
                deason_et_al2014, Starkenburg2019,
                fengqing_yu_et_al2024}.
  On the contrary, blue straggler stars can be found in almost all
  types of populations
  \cite[e.g.][]{a_sandage1953, Yanny_2000,
                ricardo_carrera_et_al2002,
                guillaume_f_thomas_et_al2019,
                belokurov_et_al2020_mnras_496_1922b},
  including the halo environment
  \cite[e.g.][]{Deason2011, deason_et_al2014, Thomas2018Atype,
                Starkenburg2019, barbosa_et_al2022,
                fengqing_yu_et_al2024, manuel_bayer_et_al2025}.
  As blue stragglers are intrinsically fainter than BHB stars (see
  e.g. figure 1 in \citealt{a_sandage1953} and
  \citealt{renzini_and_fusi_pecci1988} and figure 3
  in \citealt{gaia_collab_babusiaux_et_al2018}), a distorted
  depiction of the halo can result from imperfect cleaning of these
  contaminants. The physical difference between them is that blue
  stragglers originate from hydrogen-burning stars on the main
  sequences \citep{c_d_bailyn1995}, whereas BHB stars are giant
  helium-burning stars \citep{carroll_and_ostlie2007}. As a
  consequence, blue stragglers are more compact than BHB stars, with
  surface gravities of log$_{10}(g)$ = 2.8--3.75 and
  log$_{10}(g)$ = 3.75--5.0, respectively (see figure 7 in
  \citealt{ronald_wilhelm_et_al1999_aj117_2308w} and
  \citealt{Vickers_2012}).

\begin{figure}[h]
      \centering
      \includegraphics{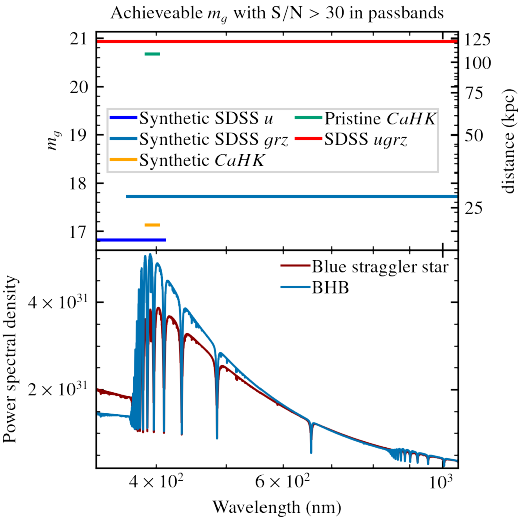}
      \caption{Wavelength dependence of faintest limit for BHB stars.
               Top panel: x axis shows the wavelength covered by 
               any of the filter passbands used in this study. For BHB 
               stars in the vetted sample from 
               \citet{bystroem_et_al2025}, the y axis then gives their 
               faintest limit (maximum apparent magnitude in the SDSS 
               \textit{g} band) where the sample still fulfils a 
               S/N greater than 30 in any of the 
               tested passbands. A second y axis shows the 
               corresponding heliocentric distance for this apparent 
               g magnitude assuming a typical BHB star with
               $M_{g_\text{SDSS}}$ = 0.5. Bottom panel: As a reference,
               we illustrate \citet{munari_et_al2005library} stellar
               library synthetic stellar spectra generated for stars
               with $T_{\text{eff}} = 8500 K$,
               $\text{bolometric luminosity}\ L%
               = 40\,\text{solar luminosity}$,
               $[\text{Z}/\text{H}] \simeq -0.5$ in the lower panel.
               One of them has log$_{10}$($g$) = 3, which is typical for
               a BHB star (the blue curve). The
               other has log$_{10}$($g$) = 5, which is typical for a
               blue straggler (the red curve).}
      \label{comparison_figure}
  \end{figure}

  In many previous studies, methods based on broad-band photometry
  have been used to discriminate BHB stars from the contaminant blue
  straggler population based on their band-colour differences (see
  e.g. \citealt{Yanny_2000}, \citealt{Deason2011},
  \citealt{deason_et_al2014}, \citealt{Fukushima_2018},
  \citealt{Thomas2018Atype}, \citealt{fengqing_yu_et_al2024}). This
  has led to BHB selections with an average completeness as low as
  50\% and contamination rates as high as 40\%. In the work of
  \citet{Starkenburg2019} these success rates are improved by
  combining \textit{CaHK} narrow-band photometry from the Pristine
  survey \citep{Starkenburg_2017} together with SDSS \textit{ugr}
  photometry, to create a sample of BHB stars with a purity of 93\%
  and a completeness of 91\%. This success can be explained by the
  narrow wavelength window of the Pristine \textit{CaHK} band, which
  coincides with the wavelength range where the flux difference
  between the BHB and blue straggler stars is large. This is
  illustrated in Figure \ref{comparison_figure}, where the bottom
  panel shows two synthetic stellar spectra with typical stellar
  parameters for a BHB star and a blue straggler. Other noteworthy
  differences are the Balmer lines, which are wider for blue
  stragglers due to pressure broadening
  \cite[see e.g.][]{clewley_et_al2002, xue2008, Xue_2011},
  and the Balmer jump (near $3645$ \AA), which is steeper for BHB
  stars \cite[e.g.][]{lenz_et_al1998, Jordi_2006, 
                      david_f_gray2005stellar_photosphere}.

  Whereas the Pristine \textit{CaHK} band is limited to the Pristine
  survey footprint, this region of wavelength space is also covered
  by \textit{Gaia} XP spectra. These are very low-resolution spectra
  ($\lambda/ \delta \lambda \approx 25\text{--}100$,
  \citealt{gaiadr3summary, de_angeli_et_al2023,
           montegriffo_et_al2023}%
  ),
  measured within the wavelength range of \textit{Gaia}'s
  \textit{BP} and \textit{RP} colour bands and sharing the all-sky
  footprint of the \textit{Gaia} satellite data. Due to their broad
  wavelength coverage and excellent calibration, XP spectra offer
  the ability to obtain synthetic photometry in many bands 
  \citep{GDR3_GSPC}, including also in SDSS \textit{ugr} and 
  \textit{CaHK} (\pristine\ \citeyear{PristineDR1}). Currently, such
  spectra are available for about 220 million sources of the
  $\sim 2 $ billion sources in \textit{Gaia} Data Release (DR) 3
  \citep{gaiadr3summary} and are shallower in their magnitude
  ($G < 17.5$) range than the deep Pristine data.

  In this work, we build on the success of the work of
  \citet{Starkenburg2019} and construct BHB catalogues using
  $u - CaHK$ and $g-r$ colour\footnote{For the remainder of this
                                       work, \textit{ugriz} bands
                                       refer to the SDSS bands. If
                                       combinations with other
                                       systems are made, such as
                                       Pan-STARRS1, this is
                                       explicitly mentioned.}
  spaces to investigate halo substructures using 3D, 4D (including
  line-of-sight velocities), and (where available) full 6D phase-space
  samples of these bright tracers. For the $CaHK$ narrow-band
  information, which leverages such improved discriminating power
  between BHB and blue straggler stars, we use two different sources,
  resulting in two catalogues: 1. the synthetic \textit{CaHK} from the
  Pristine survey DR 1 (table 1 in \pristine\ \citeyear{PristineDR1})
  that was calculated from \textit{Gaia}'s DR3 XP spectra and 2. the
  deeper \textit{CaHK} from the internal DR2.1 of the Pristine survey
  (Yuan et al., in prep.). The Pristine \textit{CaHK} photometry was
  calibrated using the synthetic \textit{CaHK} photometry. But we note
  that the dominant types of stars used for this calibration are
  FGK stars and not the A-type stars we investigate in this work. Both
  catalogues use SDSS $ugr$ broad-band information. The difference in
  depth for the various input photometric bands, and the impact these
  limits have on the resulting distances for a typical BHB star tracer,
  are illustrated in \autoref{comparison_figure}.

  While the first catalogue is shallow (limited to BHB stars within
  $\sim20$ kpc), it will no longer be limited by the Pristine Survey
  footprint of $8500 \ \text{deg}^2$. Instead, it can leverage the sky
  area covered by SDSS \cite[14555 deg$^2$,][]{alam_et_al2015_sdss_iii}.
  We refer to it as the shallow catalogue for the remainder of this
  work. The second internal DR2.1 catalogue has a narrower footprint,
  but allows us to probe distances between $\sim20$ and $\sim120$ kpc
  (Yuan et al., in prep.). It includes the footprint already covered by
  \citet{Starkenburg2019}, but extends significantly beyond it due to
  the growing footprint of the ongoing Pristine survey on the sky.
  Additionally, improvements have since been made in data reduction for
  the $CaHK$ magnitudes (see for details \pristine\
  \citeyear{PristineDR1}). We refer to this catalogue as the deep
  catalogue.

  The paper is structured as follows. In \autoref{data_section} the
  relevant datasets are described. In \autoref{selection} we outline
  our methodology resulting in two samples of (candidate) BHB
  stars presented in \autoref{samples}. We use these two datasets to
  explore various halo substructures in \autoref{science}. Section
  \ref{discussion} includes a discussion of our findings and in
  \autoref{conclusion} we present our main conclusions and the future
  outlook.

  \section{Data}\label{data_section}

  To distinguish BHB stars from blue stragglers (and other less
  significant contaminating sources), we combined the narrow-band $CaHK$
  photometry with broad-band $ugr$. In the subsections below, we
  detail each of these in turn, and discuss cleaning steps that have
  been applied to the data.

  \subsection{Sloan \textit{ugr} photometry}%
  \label{sloan_photometry_section}

  As a first step, we selected A-type stars with good-quality photometry
  from SDSS. For this purpose, we queried SDSS DR19
  \citep{sdss_data_release19}, but we note that the final SDSS imaging
  was released in SDSS DR12 \citep{alam_et_al2015_sdss_iii} and since
  SDSS DR13 \citep{albareti_et_al2017_sdss_iv} the photometry and
  object IDs have not changed. The criteria applied are the same as
  outlined in \citet{Starkenburg2019} and these are:
   \begin{itemize}
    \item primary objects ($\text{mode} = 1$) classified as stars
    ($\text{type} = 6$).
    \item $0.9 < (u-g)_0 < 1.4$ and $-0.25 < (g-r)_0 < -0.05$, where we
    use the dust reddening estimates that are also in the PhotoObjAll
    table and are based on \citet{Schlegel+98}\footnote{%
      The same applies to the \textit{CaHK} photometry, where we assume
      $A_{CaHK}/E(B\! -\! V) = 3.918$ (\pristine\
      \citeyear{PristineDR1}).}.
    \item clean photometry ($\text{clean} = 1$).
    \item unsatured \textit{ugr} photometry with no de-blending and
    interpolation issues, not close to the edge of the frame
    (see for the exact SQL query used \autoref{sql_query}).
   \end{itemize}
   These selections result in a sample of 335,247 A-type stars.

  \subsection{Removal of variable sources}%
  \label{section_variable_sources}

  Variable sources can be problematic in colour-colour spaces,
  especially when the underlying data constructing the colour-colour
  space is observed at different times. In this study, we combine data
  observed by SDSS, with that of Pristine and \textit{Gaia}, which
  together span decades of observations.

  Following \citet{Starkenburg2019}, we made use of the HERN\_VARIABLE
  value of the objInfoFlag parameter in the Panoramic Survey Telescope
  and Rapid Response System 1
  \cite[Pan-STARRS1,][]{chambers_et_al_arxiv_pan_starrs1} DR2 to
  eliminate obviously variable stars and background quasars as detected
  by the repeated observations in the Pan-STARRS1 survey. This
  deselected 4.4\% of the sample.

  Since the publication of \citet{Starkenburg2019}, many
  \textit{Gaia}-based metrics of variability have been explored (e.g.
  \citealt{belokurov_et_al2017, belokurov_et_al2020_mnras_496_1922b};
  \pristine\ \citeyear{PristineDR1}). A useful quantity is
  g$_\text{var} = \sqrt{N_{\text{obs},G}}\frac{\delta F_G}{F_G}$, where
  $N_{\text{obs},G}$ is the number of observations that
  contribute to the photometric $G$-band value and
  $\frac{\delta F_G}{F_G}$ is the relative \textit{G} flux uncertainty
  \citep{belokurov_et_al2017}. By modelling the distribution of
  g$_\text{var}$ at a given $G$, \pristine\ (\citeyear{PristineDR1})
  defined a quantity, P$_\text{var}$, for each source in \textit{Gaia}
  DR3 that gives the probability that this source is variable in nature.
  During the exploration of the possible use of P$_\text{var}$ to
  remove variable sources from our initial sample of A-type stars, we
  found that the recommended variability cut of P$_\text{var} < 0.3$ by
  \pristine\ (\citeyear{PristineDR1}) would remove many
  spectroscopically validated BHB stars. The results of these tests are
  presented in \autoref{variability_appendix_section}. This is an
  interesting result in itself, as it highlights that these stars
  intrinsically have variations at a level beyond that of (more
  standard and numerous) FGK stars in the \textit{Gaia} photometry. We
  note that \citealt{belokurov_et_al2020_mnras_496_1922b} reported
  already what they saw as an elevated binary fraction with unresolved
  stellar companions for the nearby blue straggler and blue horizontal
  branch stars. We furthermore note here that the photometric
  uncertainties at the blue end are higher in \textit{Gaia}
  \cite[e.g.][]{gedr3, riello_et_al2021}, which could be connected to
  this. For the purpose of this work, this test means we refrain from
  any more stringent cuts on variability (besides the HERN\_VARIABLE
  flag) in our sample so as to not remove bona fide BHB stars.

  \subsection{\textit{CaHK} photometry from the Pristine-%
              \textit{Gaia} synthetic catalogue}%
  \label{pristine_gaia_section}

\begin{figure*}
    \includegraphics{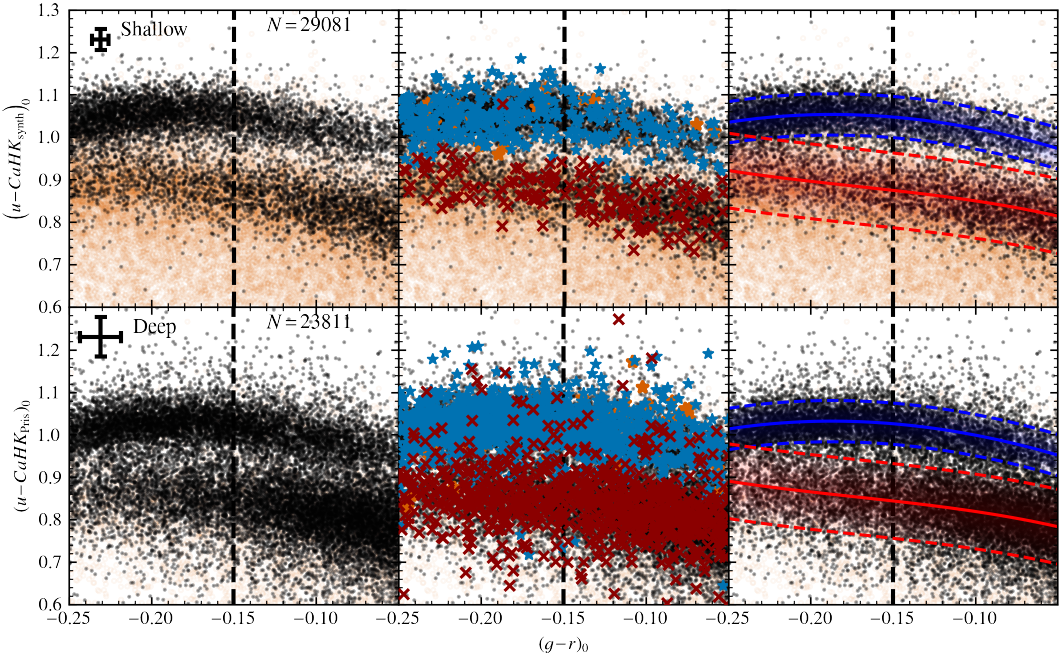}
    \caption{Left panels: $\left(u-CaHK\right) _ 0$-$(g-r) _ 0$
             colour-colour diagrams of A-type stars split into
             objects that have $E\left(B-V\right) < 0.1$ (filled and
             dark symbols) and $E\left(B-V\right)\geq 0.1$ (open and
             orangish markers). The number of objects is given in the
             top right as well as the average uncertainty (markers
             in the top left of the panel). Middle panels: Same
             as the left panels, overplotted with the training
             samples of blue straggler stars from \citet{xue2008}
             (red crosses) and the value-added catalogue of BHB
             stars in the Dark Energy Spectroscopic Instrument
             \citep{bystroem_et_al2025} (blue and orange stars for
             $E\left(B-V\right) < 0.1$ and $\geq$0.1, respectively).
             Right panels: Same as the left panels, overplotted
             with the results of the modelling curves (see text for
             details). The difference between the top and bottom panels
             is the origin of the $CaHK$ filter. In the top panels, it
             is the synthetic photometry in the Pristine \textit{CaHK}
             band from the Pristine-\textit{Gaia} synthetic
             catalogue (\pristine\ \citeyear{PristineDR1}) and in
             the bottom panel it is the Pristine survey $CaHK$.}
  \label{colour_colour_diagrams}
  \end{figure*}

  Using the official \textit{Gaia} (Early) DR3 match to SDSS
  \citep{gedr3, gaiadr3summary} to obtain Gaia IDs for the SDSS
  catalogue of A stars described in the previous section, we matched
  to the Pristine-\textit{Gaia} synthetic catalogue described in
  \pristine\ (\citeyear{PristineDR1}). Based on recommendations in
  \citet{Starkenburg2019} and \citet{GDR3_GSPC}, we implemented three
  additional quality cuts to the resulting dataset:
  \begin{itemize}
    \item $0.6 < \left(u-CaHK_\text{synth}\right)_0 < 1.3$ to
    select the colour space for which clean BHB star separation is
    possible \cite[see][]{Starkenburg2019}. This leaves 110,335
    stars.
    \item
    $\text{renormalised unit weight error}\ (\text{RUWE}) < 1.4$.
    This rather standard RUWE cut has been recommended to remove
    clear binary stars \cite[e.g.][]{gedr3, fabricius_et_al2021} and
    was also used in \citet{guillaume_f_thomas_and_battaglia2022}
    for BHB stars. This subsequent selection leaves us with 104,803
    stars.
    \item $CaHK$ signal-to-noise ratio (S/N) above 30 based on the
    recommendation in \citet{GDR3_GSPC}. This quality cut removes
    many fainter sources, as the $CaHK$ passband is at the blue end
    of the XP spectrum where the S/N is generally lower.
    After this final cut, we are left with a sample of 29,081 stars.
  \end{itemize}
  We relaxed the recommended criterion $E\left(B-V\right) < 0.3$ in
  \pristine\ \citeyear{PristineDR1} and Yuan et al. (in prep.) to avoid
  highly extincted regions and decided to not apply this criterion in
  the data compilation. While we have a more complete sample without
  this cut, we probably are more impure in the selection of BHB
  candidates since we include objects that may have unreliable
  photometry due to the presence of significant extinction. We note
  therefore that we include a $E\left(B-V\right) < 0.1$ cut for the
  science results (see \autoref{results_section}). However, the
  inclusion of all sources for the catalogues (that are made publicly
  available with this work) will allow users to make their own choices
  about the strictness of their selection dependent on their science
  case. The top left panel of \autoref{colour_colour_diagrams}
  illustrates the distribution in $(g-r) _ 0$ and
  $\left(u-CaHK_\text{synth}\right) _ 0$ for the resulting sample.

  We calculated the colour index uncertainties through Gaussian
  uncertainty propagation of the magnitude uncertainties (their
  averages are visible from the marker in the top left corner of the
  left panels in \autoref{colour_colour_diagrams}). We note that the
  average uncertainty in $(g-r) _ 0$ is 0.005, which is considerably
  smaller than the average uncertainty in
  $\left(u-CaHK_\text{synth}\right) _ 0$ of 0.03.

  From the distribution in colour index space in the top left panel
  of \autoref{colour_colour_diagrams}, it is very clear that the
  stars cluster along two separated sequences. These match the
  sequences identified in \citet{Starkenburg2019} as belonging to
  BHB stars and blue stragglers (here from top to bottom). The
  majority of stars in these two sequences have
  $E\left(B-V\right) < 0.1$ and therefore have very reliable
  photometry. The two sequences are well separated, although they
  seem to partly overlap at bluer $(g-r) _ 0$. There is also an
  accumulation of sources with very blue
  $\left(u-CaHK_\text{synth}\right) _ 0$, which belong to neither
  the distribution of blue straggler candidates nor BHB candidates.
  These stars might be young, hot main-sequence stars that we
  preferentially pick up in the parts of SDSS that are closer to the
  Galaxy's disc. Since these stars have similar surface gravities to
  blue stragglers or even higher ones
  \cite[e.g.][]{c_w_allen1973, mihalas_and_binney1981,
                sparke_and_gallagher2007,
                carroll_and_ostlie2007},
  it results in an equally weak, or weaker Balmer jump
  \cite[e.g.][]{lenz_et_al1998, Jordi_2006,
                david_f_gray2005stellar_photosphere},
  and thus a bluer colour in this space. We indeed confirm that if we
  split the samples by Galactic latitude or $E\left(B-V\right)$, the
  lower latitude or higher $E\left(B-V\right)$ samples show a very
  significantly higher fraction of these stars and that they disappear
  almost completely at higher latitudes or lower $E\left(B-V\right)$.

  \subsubsection*{Note on the use of synthetic \textit{u}-band
                  photometry}%
  \label{pristine_gaia_uband_section}
  As we are using the synthetic XP-based $CaHK$ photometry for the
  stars studied in this section, we have considered also using the
  synthetic \textit{ugr}-band photometry from the XP spectrum,
  thereby making full use of the all-sky footprint of the
  \textit{Gaia} satellite. However, in particular for the $u$-band
  data, this would result in a significant loss in depth (see also
  \autoref{comparison_figure}), as well as completeness. The
  transmission curve of \textit{Gaia BP} is low and highly
  structured in the wavelength range that covers \textit{u} and
  information in the spectra vanishes below 3300\,\AA\
  \citep[see Section 2.2.2 of][together with their figure 5]{%
    GDR3_GSPC}.
  While \citet{GDR3_GSPC} release synthetic photometry in \textit{u}
  in their \textit{Gaia} Synthetic Photometric Catalogue (GSPC), the
  recommendation is to consider as good-quality
  measurements only those with a S/N above 30. In DR3 fewer than 17\%
  of the close to 220 million sources with XP spectra have a \textit{u}
  synthetic photometry S/N above 30. For our purposes, we
  therefore do not use the \textit{Gaia} XP synthetic $u$-band
  information, but rely on the SDSS measured $u$-band data instead.
  By avoiding the troublesome synthetic SDSS \textit{u} photometry,
  we can select significantly more and/or fainter BHB stars within
  the SDSS coverage than would be possible with
  $\text{S}/\text{N} > 30$ in the GSPC.

  For studies that do utilise the synthetic $u$ band from Gaia XP
  spectra to create BHB samples, we refer the reader instead to
  \citet{wenbo_wu_et_al2025}, and \citet{guozhen_hu_et_al2025}. The
  latter work uses synthetic photometry from the SkyMapper system (a
  somewhat different $u$ band than from SDSS in terms of wavelength
  coverage).

  \subsection{\textit{CaHK} photometry from the Pristine survey}%
  \label{pristine_cahk}

  For the deep dataset that relies on ground-based Canada France
  Hawaii Telescope (CFHT) $CaHK$ photometry from the Pristine survey, we
  proceeded similarly to the manner outlined in the previous section.
  After cross-matching the catalogue of A-type SDSS stars to the
  internal DR 2.1 of the Pristine survey photometric catalogue
  (26,134,735 sources in observations until February 2026), we
  implemented the same colour index cut of
  $0.6 < (u_\text{SDSS}-CaHK_\text{Pris})_0 < 1.3$ (24,498 sources),
  and applied the RUWE $< 1.4$ cut (23,901 sources).

  While the photometry is much deeper and provides good S/N to the full
  depth of the \textit{Gaia} catalogue (so no S/N cut is necessary), we
  have to add one additional quality cut based on the photometric
  classification parameter from the Cambridge Astronomy Survey Unit 
  \cite[CASU,][]{mike_irwin_and_j_lewis2001, mike_irwin_et_al2004}
  pipeline in the Pristine survey photometric catalogue, and especially
  the combined CASU parameter as described in \pristine\
  (\citeyear{PristineDR1}), which combines the values for individual
  exposures. While in principle we would like to include only sources
  in the Pristine survey photometric catalogue that have stellar-like
  point spread functions (CASU parameter values $-$2 and $-$1), sources
  with stellar-like point spread functions in at least one of the
  exposures while being classified as extended in other exposures (CASU
  parameter value of 1) become more numerous for stars with fainter
  $CaHK$. Because we do not want to miss possible faint BHB stars at
  large distances, we keep all sources in the deep dataset that have
  values of $-$2, $-$1, 1, and 2 for the CASU parameter and remove
  objects with multiple discrepant magnitudes (i.\,e., flag is $-$3).

  This leaves us with a final dataset of 23,811 stars for this deep
  dataset, of which 3,595 sources are in overlap with the shallow
  dataset based on synthetic $CaHK$ magnitudes. The lower panels of
  \autoref{colour_colour_diagrams} show the colour-colour space for
  the deeper targets. As expected from the higher Galactic latitude
  of the Pristine footprint compared to SDSS, there is a much lower
  density of (probably blue main-sequence) sources below the blue
  straggler sequence than in the top panels.

  One additional, initially surprising, observation from a
  comparison of the top and bottom panels, is that the sequences
  seem shifted slightly in their y axis values
  $\left(u-CaHK_\text{Pris}\right) _ 0$, by 0.02--0.03 mag, i.e.
  for the deeper Pristine survey data, the sequences are
  systematically bluer. A more detailed investigation is presented in
  \autoref{appendix_cahk_synthetic_pristine_magnitude_difference}. This
  is in contrast to our expectations, as the $ugr$ photometry is derived
  from the same source, and additionally the Pristine \textit{CaHK}
  photometry was calibrated using the synthetic \textit{CaHK} photometry
  (\pristine\ \citeyear{PristineDR1}). However, we note that the
  calibration is performed on the full dataset that is dominated by FGK
  stars and that the stars selected for this work are significantly
  hotter and thus bluer in nature. A small difference in the filter
  curve used for the creation of synthetic $CaHK$ magnitudes compared
  to the filter curve on the CFHT telescope used for Pristine, could
  potentially be the underlying cause for such an offset. The results
  presented in
  \autoref{appendix_cahk_synthetic_pristine_magnitude_difference} would
  be in line with such an explanation. For the purpose of this work, we
  treat the two datasets separately and, as our method automatically
  takes this offset into account when the two sequences for BHB stars
  and blue stragglers are fitted, we do not expect the offset to affect
  our results.

  \subsection{Training and test sample of candidate BHB and blue 
              stragglers stars}%
  \label{sec:training}

  We use a catalogue of 3689 blue straggler stars from \citet{xue2008}
  and a recent catalogue of 5,441 BHB stars from the value-added
  catalogue in the Dark Energy Spectroscopic Instrument
  \cite[DESI,][]{desi_edr, desi_dr1} Milky Way Survey
  \cite[MWS,][]{andrew_p_cooper_et_al2023} DR1
  \citep{koposov_et_al2024, koposov_et_al2026} compiled in
  \citet{bystroem_et_al2025} to construct training and test samples
  for our BHB selection algorithm. The catalogue of \citet{xue2008}
  is subsequently cleaned for (highly) variable sources by use of
  the HERN\_VARIABLE value in Pan-STARRS1 DR2, which reduces the
  sample to 3,153 stars. \citet{bystroem_et_al2025} removed
  (variable) quasars and RR Lyrae stars from their sample already
  and is not cleaned further.

  In our Pristine-\textit{Gaia} catalogue, there are 256 and 451
  stars in common with these training samples for blue straggler
  and BHB stars, respectively. For the deep Pristine dataset, these
  numbers are 989 (of which 15 stars were removed because they are
  also in the catalogue of BHB stars and we trust the more recent
  classification of these stars by \citealt{bystroem_et_al2025}
  more) and 1,476 (of which 58 and 159 are in both the
  Pristine-\textit{Gaia} and Pristine dataset).

  We use 85\% of each of these samples as training samples for our
  semi-supervised hierarchical Bayesian mixture modelling (see
  \autoref{derivation_of_free_parameters}). They are shown
  overplotted on our A-star catalogues in the middle panels of
  \autoref{colour_colour_diagrams}. From these panels it is
  immediately clear that the upper sequence indeed corresponds to
  BHB stars, whereas the lower sequence contains the blue straggler
  stars. There seems to be a low level of contamination of some blue
  straggler stars from the \citet{xue2008} catalogue onto the upper
  sequence. This is not greatly surprising, as the discrimination
  between blue straggler and BHB stars in their catalogues is
  performed on relatively low-resolution SDSS spectra and is less
  carefully vetted compared to the more recent BHB sample of
  \citet{bystroem_et_al2025}. The remaining 15\% of the samples, for
  which 15\% is selected randomly redwards of the average $(g-r) _ 0$
  for the sample and 15\% bluewards, are not included in the training
  phase, but are rather used as a testing dataset to cross-validate our
  results.

  We also compare the performance of photometry from several synthetic
  photometric bands (provided by the GSPC or calculated using
  \texttt{GaiaXPy}%
  \footnote{gaia-dpci.github.io/GaiaXPy-website}) on their ability to
  select BHB stars from subsets of the training sample to the
  methodology based on the $\left(u-CaHK_\text{synth}\right)_0$-%
  $(g-r) _ 0$ in \autoref{section_colour_colour_diagrams}. In sum, our
  exploration of other colour-colour diagrams of candidate A-type stars
  found no alternative that separates the tracks of the blue straggler
  and BHB stars better than the $\left(u-CaHK\right) _ 0$-$(g-r) _ 0$
  colour-colour diagram. Therefore, we only consider the
  $\left(u-CaHK\right) _ 0$-$(g-r) _ 0$ colour-colour diagram for the
  selection of BHB stars.

  \section{An empirical model to select blue horizontal branch stars}%
  \label{selection}

  \subsection{The model}\label{model}

  To determine the probability that each star in our catalogues of
  A stars is a BHB star, we modified the hierarchical Bayesian
  mixture model from \citet{fengqing_yu_et_al2024} that was originally
  developed to select BHB stars from Dark Energy Survey (DES) DR2
  \citep{t_m_c_abbott_et_al2021survey} \textit{griz} photometry.
  This mixture model is based on three underlying populations: BHB
  stars, blue straggler stars, and outliers.
  
  Both the BHB and blue straggler stars are assumed to be distributed
  normally around the average tracks along
  $\left(u-CaHK\right) _ 0$, described by
  \begin{equation}
    \left(u-CaHK\right)_{0,\text{BHB}}%
    = \sum_{k=0}^3 a_k(g-r)_0^k
    \label{bhb_third_order_polynomial_equation}
  \end{equation}
  and
  \begin{equation}
    \left(u-CaHK\right)_{0,\text{straggler}}%
    = \sum_{k=0}^3 b_k(g-r)_0^k,
    \label{blue_straggler_third_order_polynomial_equation}
  \end{equation}
  with $a _ 0$, $a _ 1$, $a _ 2$, $a _ 3$, $b _ 0$, $b _ 1$, $b _ 2$,
  and $b _3$ as parameters of the polynomial describing the average.
  The spreads of these normal distributions are
  \begin{equation}
    \sigma_\text{obs,star type}%
    = \sqrt{\delta\left(u-CaHK\right)_0^2+\sigma_\text{star type}^2}
 , \end{equation}
  where star type can be BHB or blue straggler,
  $\delta\left(u-CaHK\right) _ 0$ is the photometric uncertainty in that
  colour combination, and $\sigma_\text{BHB}$ or
  $\sigma_\text{straggler}$ are the intrinsic dispersions of the
  distributions for the two populations. In our chosen
  parametrisation, we use $\left(u-CaHK\right) _ 0$ as the colour in
  this equation \citep[following][]{Starkenburg2019} because the
  uncertainties in this colour are typically much larger compared to
  $(g-r) _ 0$ . The intrinsic dispersions are treated as free
  parameters of the model.

  The fraction of candidate outliers modelled by free parameter
  $c _ 0$ and the final free parameter, $d _ 0$, sets the overall
  fraction of BHB stars (note that these also set the fraction of blue
  straggler stars through $1 - d_0 - c_0$). We note that our choice to
  model only the overall fractions is in contrast to
  \citet{fengqing_yu_et_al2024}, where both fractions are modelled as
  linear functions with a dependence on magnitude (e.g. the outlier
  fraction is modelled as $c_0 + c_1g_{\text{DES},0}$). In our samples,
  we do not expect the outliers, nor the BHB stars, to increase
  or decrease linearly with magnitude. Outliers may include
  intrinsically brighter young and hot main-sequence stars, as well as
  very faint white dwarfs and quasars (figure 5 of \citealt{Yanny_2000}
  and the left panels of figures 5 and 8 in
  \citealt{deason_et_al2014}). When combined with the 3D distributions
  of these sources and the survey’s footprint, this could lead to
  outlier fraction relationships with apparent magnitude that are
  challenging to predict beforehand and may not follow a linear
  pattern. Additionally, while the relative numbers of candidate blue
  straggler and BHB stars have been observed to remain roughly
  consistent across heliocentric distance (see
  \citealt{Starkenburg2019}; and \citealt{Fukushima_2018} in
  their figure 5), this relationship is not well confirmed at faint
  magnitudes.

  In our adapted version of the model by \citet{fengqing_yu_et_al2024},
  the individual log-likelihood  given the model parameters
  $\pmb\theta$ for each stellar-like source $i$ in the dataset(s), are
  given by the sum of the weighted population $j$-dependent (blue
  straggler, BHB, or outlier) log-likelihoods
  \begin{equation}
    f_j\log 
    L_j\left(\left\{\left(u-CaHK\right)_{0,i}\right\},%
    \left\{(g-r)_{0,i}\right\}%
    \Big|{\pmb\theta}\right)
    \label{likelihood}
  ,\end{equation}
  where the weights are given by
  $f_j\in \left\{f_\text{outlier}=c_0,f_\text{BHB}=d_0,%
  f_\text{straggler}=1-c_0-d_0\right\}$,
  given by the marginalisation over the possible populations as
  explained above. Moreover, an additional marginalisation over all
  possible $(g-r) _ {0,i}$ given its uncertainties for each $i$ is
  required in the form of an integral of the sum of
  \autoref{likelihood}. In practice, this integral over
  $(g-r)_{0,i}$ for each $i$ is performed in two steps. First, on a
  grid of 50 points linearly spaced between 
  $\pm 5\delta (g-r)_{0,i}$ the integral over
  $\left(u-CaHK\right) _ {0,i}$ is computed given each value of
  $(g-r) _ 0$. Second, we compute the resulting one-dimensional
  integral using a Riemannian sum over the associated
  $(g-r) _ 0$ likelihood.

  To get the log posterior probability distribution we assume uniform
  prior probability distributions for all the twelve parameters with
  limits as specified in \autoref{model_parameters}. For the limits
  connected to $a _ 0$, $a _ 1$, $a _ 2$, $a _ 3$,
  $\sigma _ \text{BHB}$, $b _ 0$, $b _ 1$, $b _ 2$, $b _ 3$, and
  $\sigma _ \text{straggler}$ we chose initial values based on the
  results of \citet{Starkenburg2019} who use the same colour-colour
  space. We subsequently increased the ranges to ensure there were no
  edge effects in the model returns. To explore the parameter space
  spanned by the multidimensional posterior probability distribution we
  follow \citet{fengqing_yu_et_al2024} and use a dynamic nested
  sampling \citep{skilling2004, skilling2006, higson_et_al2019} and
  multiple bounding ellipsoids for the prior probability distribution
  \citep{feroz_et_al2009} as implemented in \texttt{dynesty}
  \citep{speagle2020} version 2.1.3
  \citep{koposov_et_al2023dynesty_version213}.

  \begin{table}
    \caption{Definitions of uniform prior probability distributions,
             best-sampled results with the highest likelihood, and
             uncertainties of the twelve parameters of the hierarchical
             Bayesian mixture model of the
             $\left(u-CaHK\right) _ 0$-$(g-r) _ 0$
             colour-colour diagram given in the left column.}
    \label{model_parameters}
    \centering
    \begin{tabular}{llll}
      \hline\hline
      Parameter & Limit & Best estimate\\
      & & Shallow & Deep \\
      \hline\hline
      $a _ 0$ & ($-$1, 1) & 0.9080 & 0.8864\\
      $a _ 1$ & ($-$2, 2) & $-$1.4885 &  $-$1.4885\\
      $a _ 2$ & ($-$10, 10) & $-$3.4290 & $-$3.4290 \\
      $a _ 3$ & ($-$19, 19) & 1.9665 & 1.9665\\
      $\sigma _ \text{BHB}$ & (0, 0.05) & 0.0243 & 0.0243\\
      \hline
      $b _ 0$ & ($-$1, 1) & 0.7599 & 0.7292\\
      $b _ 1$ & ($-$2, 2) & $-$1.3183 & $-$1.3183\\
      $b _ 2$ & ($-$10, 10) & $-$5.1818 & $-$5.1818\\
      $b _ 3$ & ($-$19, 19) & $-$9.9668 & $-$9.9668\\
      $\sigma _ \text{straggler}$ & (0, 0.05) & 0.0440
      & 0.0440\\
      \hline
      $c _ 0$ & (0, 1) & 0.1020 & 0.0052\\
      $d _ 0$ & (0, 1) & 0.2623 & 0.3828\\
      \hline\hline
    \end{tabular}
  \end{table}

  \subsection{Derivation of the free parameters}%
  \label{derivation_of_free_parameters}

  To derive the free parameters $a _ 0$, $a _ 1$, $a _ 2$, $a _ 3$,
  $b _ 0$, $b _ 1$, $b _ 2$, $b _ 3$, $\sigma _ \text{BHB}$,
  and $\sigma _ \text{straggler}$  of the model, we utilised the
  spectroscopically vetted training sets described in
  \autoref{sec:training}. We refrained from deriving these parameters on
  the full dataset (in contrast to the approach followed in
  \citealt{fengqing_yu_et_al2024}), as tests showed that the high level
  of (probable) main-sequence stars can influence the fitting
  parameters, a problem that is avoided by using the training sets for
  the fitting. We first derived these parameters for the colour-space of
  $\left(u-CaHK_\text{Pris}\right) _ 0$-$(g-r) _ 0$ (bottom panels
  of \autoref{colour_colour_diagrams}). Secondly, because the shapes of
  the third-degree polynomials were found to be robust between the two
  different $CaHK$ colour spaces, we fixed the values of $a _ 1$,
  $a _ 2$, $a _ 3$, $b _ 1$, $b _ 2$, and $b _ 3$,
  $\sigma _ \text{BHB}$, and $\sigma _ \text{straggler}$ and only
  refitted $a _ 0$ and $b _ 0$ using the shallow training sample,
  thereby taking care of the offset between the two $CaHK$ photometry
  sources (see discussion in \autoref{pristine_cahk} and
  \autoref{appendix_cahk_synthetic_pristine_magnitude_difference}). We
  note that we kept $\sigma _ \text{BHB}$ and
  $\sigma _ \text{straggler}$ fixed as the intrinsic widths of the
  sequences should be maintained in the two spaces (even though the
  sequences might show different widths in practice, as the
  observational uncertainties vary between the datasets). We find that
  $\sigma _ \text{straggler} > \sigma _ \text{BHB}$, as also found in
  \citet{Starkenburg2019}.
 
  The results of the initial sampling on the training sample deep
  dataset are well behaved within the prior ranges
  (see \autoref{corner_plot_training} and
  \autoref{corner_plot_training_synthetic_version} in
  \autoref{additional_corner_plots} for the deep and shallow
  dataset respectively). As expected, since these parameters
  collectively describe
  the positions of the BHB and blue straggler populations, there is a
  positive correlation among $a _ 0$, $a _ 1$, and $a _ 2$, and $b _ 0$,
  $b _ 1$, and $b _ 2$ (the Spearman rank correlation coefficients
  range between 0.30 and 0.78), and moreover a correlation between $c_0$
  and the intrinsic dispersions given to the sequences
  ($\sigma _ \text{BHB}$, and $\sigma _ \text{straggler}$), again to be
  expected. We furthermore find marginal correlations between some of
  the other parameters, but of much smaller amplitude in their Spearman
  rank correlations. The right panels of
  \autoref{colour_colour_diagrams} show the fitted third-degree
  polynomials and the associated intrinsic standard deviations. The
  models confirm the better separation of the populations of BHB and
  blue straggler candidates for redder colours (right) of the
  $(g-r) _ 0$ range compared to the blue (left), where the polynomials
  touch each other within the $2\sigma$ range.
  
  Finally, we fix the fitted parameter values for each of the training
  sample runs and use these for the respective full dataset runs to
  converge on the final values for $c _ 0$ and $d _ 0$ that represent
  the fractions of stars in the data samples that can be classified as
  BHB stars and as outliers (see
  \autoref{corner_plot_synthetic_version} and \autoref{corner_plot} in
  \autoref{additional_corner_plots} for the posterior distributions).
  All final values for the parameters are presented in
  \autoref{model_parameters}.
      
  \subsection{Two catalogues of candidate blue horizontal branch stars}%
  \label{samples}
  
  To calculate the probability that a source in one of our two samples
  is either a candidate BHB star, blue straggler star, or outlier, the
  posterior probability of a candidate star belonging to one of these
  three classes based on their position in colour space is given by

 \begin{equation}
    P_\text{BHB}%
    = \frac{d_0 L_\text{BHB}}%
    {d_0 L_\text{BHB}+(1-c_0-d_0)L_\text{straggler}%
     +c_0 L_\text{outlier}},
    \label{posterior_probability_bhb}
  \end{equation} 
  \begin{equation}
    P_\text{straggler}%
    = \frac{(1-c_0-d_0) L_\text{straggler}}%
    {d_0 L_\text{BHB}+(1-c_0-d_0)L_\text{straggler}%
     +c_0 L_\text{outlier}},
    \label{posterior_probability_blue_straggler}
  \end{equation}
  and
  \begin{equation}
    P_\text{outlier}%
    = \frac{c_0 L_\text{outlier}}%
    {d_0 L_\text{BHB}+(1-c_0-d_0)L_\text{straggler}%
     +c_0 L_\text{outlier}},
    \label{posterior_probability_outlier}
  \end{equation}
  respectively.
  
\begin{figure}
    \includegraphics{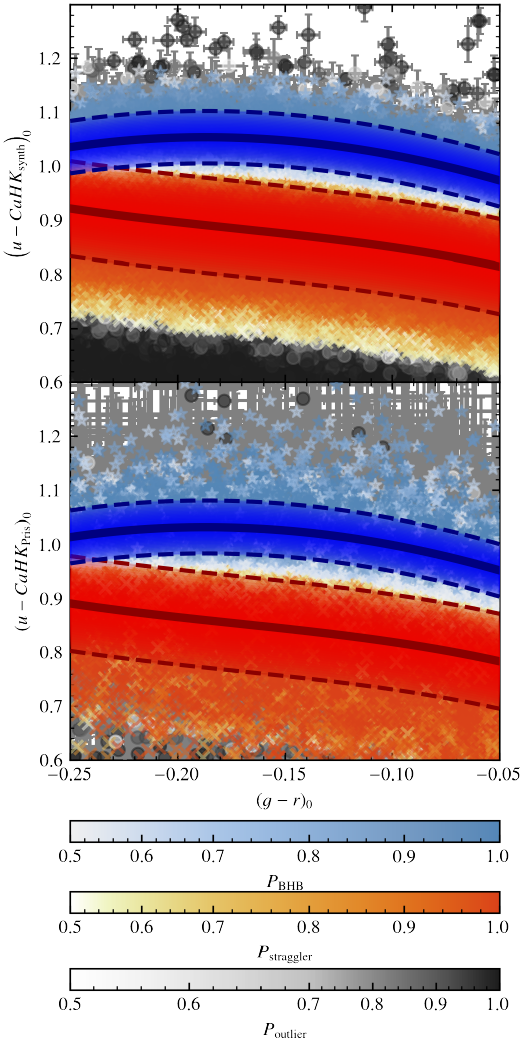}
    \caption{Stars classified as BHB, blue straggler, and outlier
             candidates are represented in blue stars, red xs, and
             grey circles, respectively. The predicted class is
             determined by the highest posterior-marginalised
             probability, and the colour of each star corresponds to
             this probability. The solid blue and red lines
             represent the third-degree polynomials
             $\left(u-CaHK\right) _ {0,\text{BHB}}$ and
             $\left(u-CaHK\right) _ {0,\text{straggler}}$,
             calculated from the best-fit parameters in
             \autoref{model_parameters}. The shaded regions with
             dashed lines along their boundaries indicate twice the
             intrinsic scatter $\sigma _ \text{obs,star type}$ for each
             sequence.}
    \label{object_plot_with_their_probabilities}
  \end{figure}

  In \autoref{object_plot_with_their_probabilities}, we present a
  colour-colour space visualisation of the predicted class
  probabilities for all the stars. The predicted class is determined
  by the highest posterior-marginalised probability. We note that
  the classification does not necessarily follow the lines of the
  third-degree polynomials and their 2$\sigma$ lines, as also the
  observational uncertainties for each star are taken into account
  in the model and the fractions of BHB and blue straggler
  candidates are different. This is most noticeable in the lower
  regions of the panels where stars are still classified as blue
  straggler below the 2$\sigma$ blue straggler fitted line. In
  general, we do find more stars classified as blue stragglers than
  as BHB stars, in line with other photometric studies in the
  literature
  \cite[e.g.][]{george_w_preston_et_al1994, Yanny_2000, Deason2011,
                 deason_et_al2014, santucci_et_al2015a}.
  Between the two datasets, we classify a significantly larger
  number of stars as outliers in the shallow dataset than in the
  deep dataset and they are preferentially clustered at low/blue
  $\left(u-CaHK_\text{synth}\right) _ 0$. As described in
  \autoref{pristine_gaia_section}, these outliers could be young and
  hot main-sequence stars that are preferentially picked up in the
  shallow dataset as it is both more nearby and has more fields
  close to the Milky Way disc.
  
  We subsequently tested the performance of the method by determining
  the purity and completeness of our selection of candidate BHB and
  blue straggler stars compared to the test samples described in
  \autoref{sec:training}. Given that we do not consider true outliers
  in the training and test samples, we can only determine upper limits
  of the purity, because we do not consider true outliers that are
  incorrectly identified as BHB or blue straggler stars in our method.
  Although we note that this might lower the real purity of the
  selection, we also stress that previous studies have shown that the
  outlier fraction is very low
  \citep[typically of the order of a few per cent,][]{%
    deason_et_al2014, deason_et_al2018astrophysical_journal,
    Fukushima_2018, amarante_et_al2024, fengqing_yu_et_al2024}.
  Additionally, we validated the samples of BHB candidates by
  comparing them to the sample of BHB stars from
  \citet{barbosa_et_al2022}. This extra
  validation sample was selected to have 100\% probability of being a
  BHB star according to these authors and is not included in the DESI
  MWS DR 1 catalogue to avoid overlap with our training and test
  samples from \citet{bystroem_et_al2025}. This adds 824 and 943 stars
  in the shallow and deep test datasets, respectively.
  
\begin{figure}
    \includegraphics{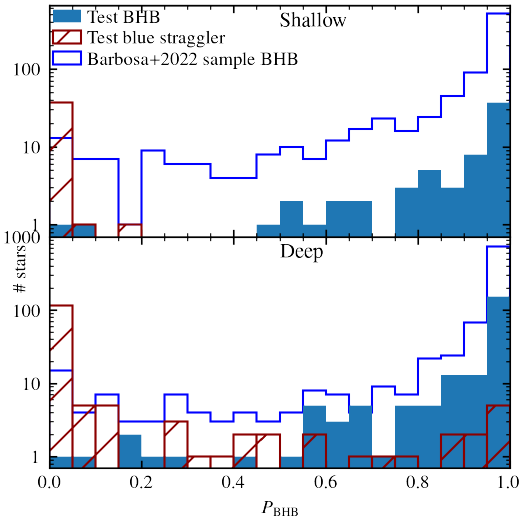}
    \caption{Probability that a BHB and blue straggler star in the
             test sample or sample of \citet{barbosa_et_al2022}
             (excluding the overlap with DESI) is a candidate BHB
             star according to the methods described in the main
             part of \autoref{selection}. The top panels show the
             results for the shallow dataset, the lower panels
             correspond to the results of the deep dataset.}
    \label{figure_bhb_posterior_probabilities}
  \end{figure}

  We used the following definitions for purity and completeness of the
  selection of BHB candidates:
  \begin{equation}
    \text{purity} =  \frac{N _ \text{BHB, BHB}}%
    {N_\text{BHB, BHB} + N_\text{BHB, straggler}}
  \end{equation} 
  and
  \begin{equation}
    \text{completeness} =  \frac{N _ \text{BHB, BHB}}{N _ \text{BHB}}
  ,\end{equation} 
  where $N _ \text{BHB, BHB}$ are the number of BHB stars from the test
  and validation samples that have been correctly identified by our
  method when we follow the maximum probability assignment of
  $\left(P_\text{BHB},\ P_\text{straggler},\ %
  P_\text{outlier}\right)$, and
  $N_\text{BHB, straggler}$ is the number of blue straggler candidates
  that have been misclassified. $N _ \text{BHB}$ is the total number of
  BHB stars in the test. The same equations can also be used for the
  blue stragglers, by switching the terms. Figure
  \ref{figure_bhb_posterior_probabilities} shows the results of these
  tests. The outcomes in percentages are given in
  \autoref{table_purity_and_completeness} and highlight the
  very high success rates of purity (an upper limit of 94\%) and
  completeness (91\% and higher) in both our BHB samples with respect
  to the test and validation samples, although an important note is
  that the test and validation samples might themselves have some level
  of impurity and incompleteness.

  \begin{table*}
    \centering
    \caption{Success rates of our BHB and blue straggler candidate 
             selections.}
    \begin{tabular}{lll}
      \hline
      Candidate sample & Purity & Completeness\\
      \hline
      Test BHBs with SDSS \textit{ugr} and synthetic \textit{CaHK} 
      & 100\% & 93\%\\
      Subset of (1) BHBs with SDSS \textit{ugr} and synthetic
      \textit{CaHK} 
      & 100\% & 92\%\\
      Test blue straggler candidates with SDSS \textit{ugr} and 
      synthetic \textit{CaHK}
      & 93\% & 100\%\\
      \hline
      Test BHBs with SDSS \textit{ugr} and Pristine \textit{CaHK} 
      & 94\% & 91\%\\
      Subset of (1) BHBs with SDSS \textit{ugr} and Pristine
      \textit{CaHK} 
      & 99\% & 94\%\\
      Test blue straggler candidates with SDSS \textit{ugr} and 
      Pristine \textit{CaHK}
      & 95\% & 91\%\\
      \hline
    \end{tabular}
    \tablefoot{The purity and completeness are upper limits.}
    \tablebib{(1)~\citet{barbosa_et_al2022}.}
    \label{table_purity_and_completeness}
  \end{table*}
  
\begin{figure}
    \includegraphics{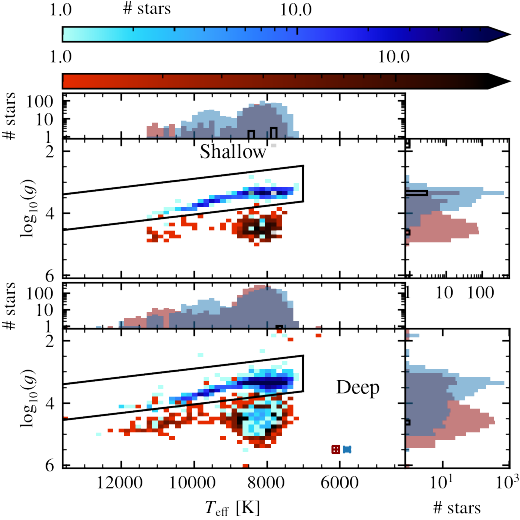}
    \caption{Two-dimensional histograms in a Kiel diagram and the 
             one-dimensional histogram projections (top and right of 
             each panel on the left and right) of all classified 
             stellar-like sources that have a effective temperature
             and surface gravity in the DESI
             \citep{desi_edr, desi_dr1} MWS 
             \citep{andrew_p_cooper_et_al2023} DR1 
             \citep{koposov_et_al2024, koposov_et_al2026} 
             from the RVSpecFit pipeline. We colour-code by their
             classification (candidate BHB stars with blue,
             candidate blue stragglers with red, and, few, candidate
             outliers with greyish colours). To the Kiel diagrams we
             show the $1.50 + 0.00014T_\text{eff}$ and
             $2.65 + 0.00014T_\text{eff}$ lines that were used in
             also a Kiel diagram by \citet{bystroem_et_al2025} to
             select BHB stars. While in the lower right corner of
             the lower panel uncertainty markers are visible, we did
             not plot for the shallow dataset since they are
             significantly smaller than the bin size.}
    \label{purity_and_completeness_figure}
  \end{figure}

  To investigate our samples across the Kiel diagram, we present in
  \autoref{purity_and_completeness_figure} the stars in our sample
  that have good-quality\footnote{We follow the quality cuts by
  \citet{bystroem_et_al2025}: RR\_SPECTYPE = STAR, remove stars that
  have the CHISQ\_WARN, RV\_WARN, and RVERR\_WARN bits set in the
  RVS\_WARN bitmask in the \texttt{RVSpecFit} pipeline, and keep
  only the primary DESI MWS DR1 observations by PRIMARY = True.}
  DESI MWS DR1 values from the \texttt{RVSpecFit} pipeline. We can clearly
  identify two sequences, with a large majority of the BHB
  stars in the shallow dataset falling in the top sequence that is
  confined within the black overplotted box following the BHB
  classification area by \citet{bystroem_et_al2025}. From the two
  panels in \autoref{purity_and_completeness_figure} we note that
  the shallow dataset is brighter and therefore also will have
  DESI spectra with a higher S/N. Indeed, we see that more
  precise stellar $T _ \text{eff}$ and log$_{10}(g)$ for the shallow
  dataset make the population of BHB and blue straggler candidates
  more discernible, with just a few BHB candidates outside of the
  BHB box region. For the deep dataset there are more BHB candidates
  that fall outside of the outlined BHB area. 94 stars (a fraction
  of 1.1\%), which we define as BHB candidates based on the
  colour-colour space are assigned higher surface gravities by DESI,
  but they typically have S/Ns of less than 20 in
  the red arm of the DESI spectrograph (see also
  \autoref{signal_to_noise_ratio}). 
    
  We also calculated local completeness and upper limits of the purity
  of the selection of BHB candidates in bins or clusters of the stars in
  the test and extra \citet{barbosa_et_al2022} validation samples in
  the colour index space making sure that each bin or cluster has a
  significant number of stars given Poisson variance. We found the
  local completeness and purity for both samples in the deep dataset to
  be lower on the order of 80\% at the blue-$(u-CaHK_\text{Pris}) _ 0$
  and red-$(g-r) _ 0$ end, while it increases to on the order of 100\%
  towards the red-$(u-CaHK_\text{Pris}) _ 0$ and blue-$(g-r) _ 0$ end.
  For the shallow dataset, there are not enough stars in the test and
  extra \citet{barbosa_et_al2022} validation samples to quantify a local
  completeness and purity in the colour index space reliably.
  
  To test our consistency between the shallow and deep samples, we
  cross-matched the stars in common and compare their outcomes. A full
  analysis can be found in \autoref{discrepant_probabilities}, but,
  in summary, we find that 97.1\% of the stars that are given a BHB
  classification in the shallow dataset, also have the same
  classification in the deep dataset and that only a very small
  subset of the stars is not consistent with its classification. For
  this small subset of stars we did not find any trend of the
  difference of the two probabilities for each classification in the
  two samples with $(g-r) _ 0$ and their
  $(u-CaHK_\text{synth}) _0$/$(u-CaHK_\text{Pris}) _0$
  uncertainties.
  
  \subsection{Distances and radial velocities}\label{distanceRV}

  To compute distances to our candidate BHB stars, we used the
  relation between the absolute \textit{g} magnitude and $(u-r) _ 0$ as
  determined by \citet{barbosa_et_al2022}:

  \begin{equation}
    \label{eq:distance}
      M_{g,\text{BHB}}%
      = \frac{0.721}{(u-r)_0} - 0.212
.  \end{equation} 

  In the work of \citet{barbosa_et_al2022} it is mentioned that the
  intrinsic uncertainty in their absolute magnitude with $(u-r) _ 0$
  trend is $\sim 0.1$. We note that the uncertainties in the colour
  $(u-r) _ 0$ is in all cases much smaller than this intrinsic
  uncertainty (at most 20\%) for the shallow dataset. A comparison with
  photogeometric, heliocentric distances from
  \citet{bailer_jones_et_al2021} corroborates the report by
  \citet{barbosa_et_al2022} (see their figure 5) who find that these
  photogeometric distances are underestimated in relation to the
  distances derived from photometry for these halo BHB stars.
  
  The shallow sample covers distances which are still within the inner
  halo (with heliocentric distances $D_\text{h}$ <22.8 kpc). The deeper
  BHB candidate sample probes much further into the halo, with
  distances reaching $\sim 120$\,kpc
  
  Where possible, we cross-matched our catalogues to obtain proper
  motion data from \textit{Gaia} DR3 and line-of-sight velocities. The
  latter are provided by \textit{Gaia} DR3, the DESI MWS DR1, the Large
  Sky Area Multi-object Fibre Spectroscopic Telescope
  \cite[LAMOST,][]{lamost} low- and mid-resolution DR11 version 2.0
  stellar and LAMOST general catalogues, the Sloan Extension for
  Galactic Understanding and Exploration (SEGUE) DR10
  \citep{ahn_et_al2014}, and DR4 \citep{buder_et_al2025data_release4}
  in Galactic Archaeology with the High Efficiency and Resolution
  Multi-Element Spectrograph
  \cite[Galactic Archaeology with HERMES, GALAH,][]{%
    sheinis_et_al2015, de_silva_et_al2015survey}.
  In case of the LAMOST low-resolution general catalogue, we calculated
  line-of-sight velocities from the tabulated redshifts. Additionally,
  we attempted cross-matches with the Radial Velocity Experiment (RAVE)
  DR6 \citep{steinmetz_et_al2020velocity_experiment} and the Apache
  Point Observatory Galactic Evolution Experiment
  \cite[APOGEE,][]{majewski_et_al2017} DR17
  \citep{sdss_data_release17}, but found no matches due to the
  different magnitude range and footprint those surveys focus on. A
  summary of the added line-of-sight velocities from each survey is
  given in \autoref{survey_line_of_sight_velocities}. The surveys are
  given priority in the given order in
  \autoref{survey_line_of_sight_velocities}; for example, if a star has
  a line-of-sight velocity in DESI and LAMOST, the DESI one is
  prioritised.
    
  \begin{table}
    \begin{center}
      \caption{Summary of the different surveys we matched our BHB
               sample(s) with to obtain line-of-sight velocities, along
               with the number of line-of-sight velocities obtained
               from each survey.}
      \begin{tabular}{lll}
        Survey & \# stars & Priority\\
        \hline
        \textit{Gaia} & 7 & 1\\
        GALAH & 9 & 2\\
        DESI & 2095 & 3\\
        SEGUE & 1811 & 4\\
        LAMOST mid-resolution stellar & 32 & 5\\
        LAMOST mid-resolution general & 150 & 6\\
        LAMOST low-resolution stellar & 1157 & 7\\
        LAMOST low-resolution general & 913 & 8\\\hline
        Total & 6172 & 
      \end{tabular}
      \tablefoot{We found no matches with APOGEE and RAVE and so these
                 are not included.}
    \end{center}
    \label{survey_line_of_sight_velocities}
  \end{table}
  
  To put all line-of-sight velocities on the same scale, we corrected
  for known systematics, such as between the LAMOST low-resolution
  spectroscopic survey and SDSS DR19. Here, we followed the LAMOST DR11
  release note%
  \footnote{\url{https://www.lamost.org/dr11/v2.0/doc/release_note}}
  and added a correction of 4.31 km s$^{-1}$ to the line-of-sight
  velocity from LAMOST low-resolution
  \cite[see also figure 2 in][]{meszaros_et_al2025}, LAMOST
  mid-resolution DR11 version 2.0, and the bright programme of the main
  part of the DESI MWS (see section 4.3 and the top panel of figure 11
  in \citealt{koposov_et_al2026}). We do not add the dispersion
  mentioned in these release notes to our uncertainties, since the
  LAMOST uncertainties are already found to be overestimated by a factor
  $\sim 2.5$ \citep{Tsantaki_2022}. The line-of-sight velocity
  differences of stars in the backup programme of the main part of the
  DESI MWS do show more significant offsets (see section 4.3 and figure
  11 in \citealt{koposov_et_al2026}) that we corrected following
  appendix section C in \citet{koposov_et_al2026}. Finally, based on
  the found zeroth order offset of $-$4.00 km\,s$^{-1}$ between SEGUE
  and the \textit{Gaia} Radial Velocity Spectrometer
  \cite[RVS,][]{cropper_et_al2018} in \citet{Tsantaki_2022} (see their
  section 6.3 and also their figure 21), we added this value to the
  SEGUE line-of-sight velocities. 
    
\begin{figure}
    \includegraphics{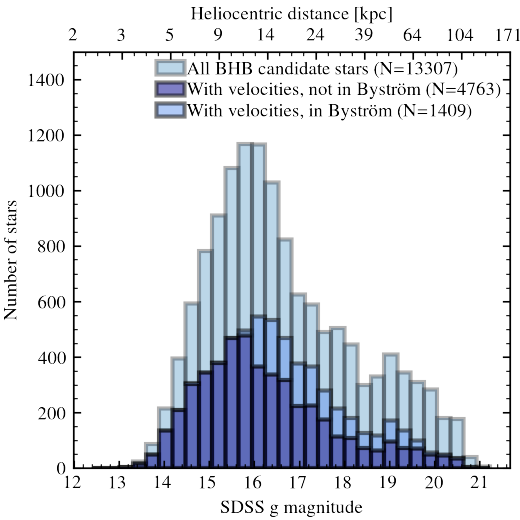}
    \caption{Comparison of our candidate BHB sample (light blue) with
             line-of-sight velocities that are not included in
             \citet{bystroem_et_al2025} (dark blue) and those that
             overlap (middle blue). The approximate heliocentric
             distance on the upper x axis is estimated utilising a fixed
             absolute magnitude of 0.5 in the g band for all BHB stars.}
    \label{bhb_candidates_vs_bystroem_histogram}
  \end{figure}

  Figure \ref{bhb_candidates_vs_bystroem_histogram} shows the resulting
  sample of BHB stars, both with and without radial velocities. To
  emphasise the additional value of this sample, we highlight the
  DESI-based sample of \citet{bystroem_et_al2025}, which currently
  comprises the most extensive collection of BHB stars. Notably, we
  expand the local population of candidate BHB stars compared to
  \citet{bystroem_et_al2025}, but we also add many stars at larger
  distances.
    
  \subsection{Final catalogues of BHB stars}%
  \label{BHBtable}
  
  As part of this work, we publish the resulting set (merging our two
  catalogues) of all stars with BHB classification according to our
  method. The final table, which is available online at the CDS
  includes the necessary Gaia identification data (Gaia DR3 ID and
  coordinates), all (de-reddened) photometric colours used, the
  resulting maximum likelihoods for the three classes (BHB, blue
  straggler, and outlier) and the inferred distances for likely BHB
  stars. We also include radial velocities if available through the
  literature cross-match (see \autoref{distanceRV}), as well as
  Galactocentric positions, angular momentum, and energies in two
  different Galactic potentials used for scientific explorations in
  \autoref{results_section}. Finally we include a flag that indicates
  if the BHB star could be identified with a known substructure, as
  presented in \autoref{results_section}. The full table with column
  names and descriptions is given in \autoref{dataset_description}.
  
  \section{Scientific exploration}%
  \label{science}
  
  This section presents the first scientific exploration of the
  resulting sample of BHB candidates obtained using the probabilistic
  classification method. Firstly, we look at several known stellar
  streams and overdensities in the Milky Way (halo), such as the
  \citet{Dodd_2023} groups, the Sagittarius stream, and the Pisces
  Overdensity. Finally, we perform an analysis to look for asymmetry in
  bulk motion between the northern and southern Galactic hemispheres.
      
  \subsection{Additional quality cuts}%
  \label{qualitycuts}
  Throughout this section, we enforce a cut of
  $P _ \text{BHB} \geq 0.5$ to select BHB candidates from the two
  samples \citep[following][]{fengqing_yu_et_al2024}. We furthermore
  apply a cut removing all sources with $E\left(B-V\right) > 0.1$ to
  avoid highly extincted regions where the de-reddening -- and thus the
  final classification of the star in the colour-colour space -- is more
  uncertain. We note that only the shallow dataset is significantly
  affected by this cut. Additionally, the photometry becomes less
  reliable in very crowded regions, such as near the centres of
  globular clusters. Therefore, we excluded the vicinity of globular
  clusters within our distance range using the 2010 version of the
  \citet{Harris1996} catalogue and eliminate BHB candidates that are
  found within a radius of one degree from their centres. This removes
  ten stars from the shallow dataset, which is 0.19\%. Finally, we
  remove from our dataset two stars that seem unbound from the Milky
  Way given their properties in the catalogue, and stars that have a
  missing or unusually high radial velocity uncertainty (see for more
  details \autoref{star_removal}).

\begin{figure}
    \includegraphics{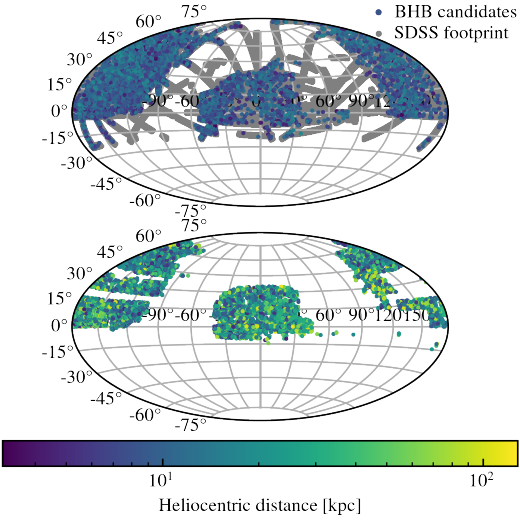}
    \caption{Three-dimensional maps of selected BHB star candidates
             based on synthetic \textit{CaHK} and SDSS IV \textit{ugr}
             photometry (top) and Pristine survey \textit{CaHK} and SDSS
             DR19 \textit{ugr} photometry (bottom), colour coded by
             heliocentric distance and avoiding highly extincted
             regions through $E\left(B-V\right) < 0.1$.}
    \label{figure_equatorial_coordinates}
  \end{figure}

  Figure \ref{figure_equatorial_coordinates} presents the
  three-dimensional (right ascension, declination, heliocentric
  distance as colour) maps for the two datasets. We caution that both
  samples are not free from selection effects as the shallow data
  suffers from varying quality due to the \textit{Gaia} scanning
  pattern and also the Pristine fields have not all been observed in
  equal quality circumstances. Visible structures thus always have to
  be treated with care and validated.
  
  \subsection{Complementing solar neighbourhood 6D substructures with
              the shallow sample}%
  \label{results_section}
  
  Dynamical quantities are a very powerful way to recover phase-mixed
  substructure \citep[see][and references therein]{Helmi_2020}. Here,
  we consider the (often used) combination of energy and the Cartesian Z
  and perpendicular components of the angular momentum
  $\left(E,\ L_Z,\ L_\perp\right)$
  to search for substructures in our BHB halo samples. The total
  energy was computed as
  \[E = \frac{1}{2}\mathbf{v}^2 + \Phi(x,\ y,\ z),\]
  where $\mathbf{v}$ and $\Phi(x,\ y,\ z)$ are the Galactocentric total
  velocity and Galactic gravitational potential at the Galactocentric
  location $(x,\ y,\ z)$ of the star. When determining the
  Galactocentric phase space coordinates, we used J2016 format right
  ascension and declination and their proper motion from \textit{Gaia}
  DR3. We corrected for solar motion, using
  $\left(U,\ V,\ W\right)_\odot%
  = (11.1,\ 12.24,\ 7.25)\,\text{km}\,\text{s}^{-1}$
  \citep{schoenrich_et_al2010}, and used
  $v_\text{LSR} = 232.8$\,km\,s$^{-1}$ for the motion of the local
  standard of rest (LSR) and $R_\odot = 8.2$\,kpc to be consistent with
  \citet{mcmillan2017}.

\begin{figure}
    \includegraphics{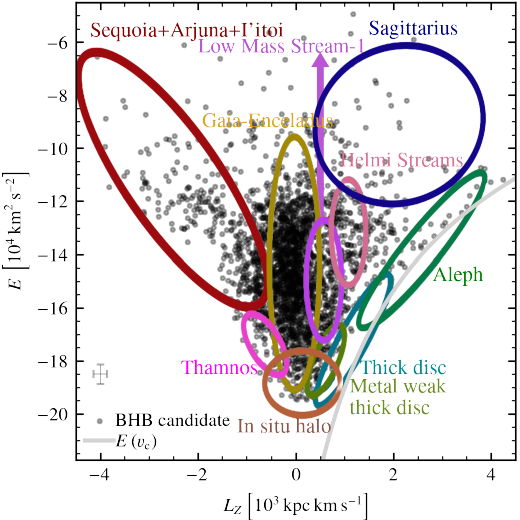}
    \caption{Energy versus angular momentum in Galactocentric,
             Cartesian $Z$ direction. The substructure regions 
             identified by \citet{Naidu_2020} using red giant stars in
             the H3 survey \citep{conroy_et_al2019h3survey}, as shown
             in figure 6 of \citet{viswanathan_et_al2025pristine}, are
             represented as coloured ellipses. These ellipses are used
             as references to associate BHB candidates with SDSS DR19
             \textit{ugr} photometry and synthetic \textit{CaHK} data
             from this work to these substructures. Additionally, as a
             reference we plot the average uncertainties with the bottom
             left marker and also the energy curve for a star on a
             circular orbit with velocity, $v _ \text{c}$.}
    \label{integrals_of_motion_results}
  \end{figure}

  Figure \ref{integrals_of_motion_results} shows the dynamical
  quantities $\left(E,\ L _ Z\right)$ for the shallow catalogue (as it
  has better quality proper motion information) with overplotted
  several ellipses that illustrate the location in this space of known
  halo substructures as identified by \citet{Naidu_2020} for their
  sample of (typically further out) H3 survey red giants
  \citep{conroy_et_al2019h3survey} and similarly shown in Figure 6 of
  \citet{viswanathan_et_al2025pristine} for their sample of halo red
  giants. For consistency, we assume $Z_\odot = 0.014$\,kpc
  \citep{binney_et_al1997} and use the same \citet{mcmillan2017}
  gravitational potential as in \citet{viswanathan_et_al2025pristine}.
  Qualitatively, we see from \autoref{integrals_of_motion_results} that
  the halo structure from our BHB stars is distinctly non-symmetric in
  ${L_Z}$, and that many of our BHB candidates are overlapping with
  known substructures. We note that higher-energy substructures such as
  debris from the Sagittarius dwarf galaxy are less populated due to
  the limited distance of the shallow sample.
    
  To quantify the substructures more robustly, we subsequently explored
  whether our BHB candidates can be linked to previously identified 6D
  (phase-mixed) substructures from \citet{Dodd_2023}. They applied a
  single-linkage-based clustering algorithm to identify overdensities
  in dynamical-quantity space by using a \textit{Gaia} DR3 sample of
  the local halo within 2.5 kpc from the Sun
  \citep[see also][for a detailed description of the clustering
                   algorithm]%
  {L_vdal_2022,Ruiz_Lara_2022}. The accurate distances that we can
  obtain for our BHB candidates, combined with radial velocities and
  proper motions, allow us to probe these substructures further out and
  monitor their extent.
  
  We closely followed the methods in \citet{Dodd_2023} by applying the
  Mahalanobis distance to their identified 27 groups for each BHB
  candidate in the three-dimensional dynamical-quantity space
  (taking into account the uncertainties of the three dynamical
  quantities and their covariances). For consistency we now use
  $Z_\odot = 0.000$\,kpc and the same gravitational potential as in
  \citet{Dodd_2023}, which consists of a \citet{miyamoto_and_nagai1975}
  model for the disc, \citet{hernquist1990model} model for the bulge,
  and Navarro-Frenk-White model \citep{navarro_et_al1997} for the halo
  with parameters $(a, b) = (6.50, 0.26)$ kpc and
  $M_\text{d} = 9.3 \times 10^{10} \text{M}_{\odot}$ for the disc,
  $r_\text{s} = 0.7$ kpc and
  $M_\text{b} = 3.0 \times 10^{10} \text{M}_{\odot}$ for
  the bulge, and $r_\text{s} = 21.5$ kpc, $\text{concentration}=12$,
  and $M_\text{halo} = 10^{12} \text{M}_{\odot}$ for the halo.
    
\begin{figure*}
    \includegraphics{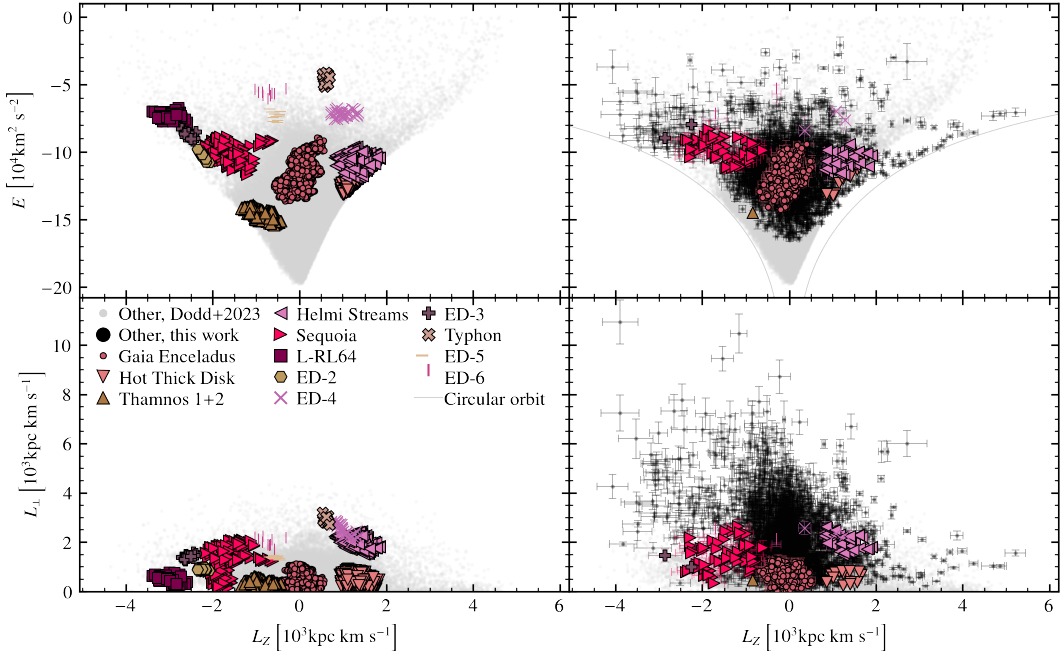}
    \caption{Dynamical-quantity space used by \citet{Dodd_2023}
             (left panels reproduced here) and here to assign BHB
             candidates in the shallow dataset to the clusters in
             \citet{Dodd_2023} (right panel). `Other' refers here to
             stars that did not get a substructure classification or
             are associated with other smaller substructures. The grey
             curve(s) on the top right-hand side is $E$ assuming the
             stars is on a circular orbit.}
    \label{results}
  \end{figure*}

  The Mahalanobis distance from a BHB candidate to a substructure
  can be viewed as a membership probability of belonging to that
  substructure. For the BHB candidates with a Mahalanobis distance
  less than 2.83
  \cite[95.4\% of each substructure falls within that distance,][]{%
    L_vdal_2022}
  we assign membership to that substructure. In \autoref{results},
  the left panels show the original representation of the
  dynamical-quantity space considered by \citet{Dodd_2023}, with
  ungrouped stars from their sample in the background. The right
  panels overlay the BHB candidates in the shallow dataset coloured
  by the group they have been found to belong to, if applicable. Out
  of the 2,769 BHB candidates, 437 can be assigned to a
  substructure:
  [218, 32, 15, 1, 44, 52, 22, 26, 10, 8, 3, 1, 1, 2, 2] in
  [Gaia-Enceladus, L-RL3, hot thick disc, Thamnos, Helmi Streams,
   Sequoia, ED-1, 13, 14, 16, ED-4, 20, 21, ED-3, ED-6]. Here, we
  would like to specify that L-RL3 \citep{L_vdal_2022}, ED-1, 13,
  14, 16, 20, and 21 are not pure substructures and are partly
  associated with other larger populations of substructure, such as
  Gaia-Enceladus \citep{Dodd_2023}. That is why we only consider
  Gaia-Enceladus, hot thick disc, Thamnos
  \citep{koppelman_et_al2019_631_l9}, Helmi streams
  \citep{helmi_et_al1999helmi_streams}, Sequoia
  \citep{myeong_et_al2019}, ED-4, ED-3, and ED-6 \citep{Dodd_2023}
  as the main groups of the substructures.
  
  In the upper right panel of \autoref{results} there is a striking
  cluster of unassigned BHB candidates at
  $L _ Z\sim -10^3\,\text{kpc}\,\text{km}\,\text{s}^{-1}$ and
  $-7\lesssim E\,10^{-4}\,\text{km}^{-2}\,\text{s}^2\lesssim -6$ that
  also clusters in $L_\perp$. These are likely BHB candidates in the
  extended part of the larger globular cluster NGC\,5466
  \citep[e.g.][]{%
    baumgardt_and_hilker2018, ferrone_et_al2023globular_clusters} 
  beyond one degree of its centre.
  
  \begin{figure}[h]
    \includegraphics{%
      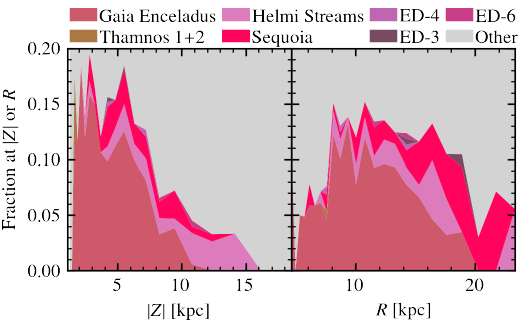}
    \caption{Left: $|Z|$ vs fraction at $|Z|$ results of the BHB
             candidates in the main groups of (re)identified
             \citet{Dodd_2023} substructure where we found BHB
             candidates. These fractions also include the rest of
             the BHB candidates that did not get a substructure
             classification or we associate to other, such as the
             hot thick disc, or smaller substructures in light grey.
             That means all the shown $|Z|$ bins sum up to a
             fraction of 100\%. The fraction is cut off at 0.2 for
             clarity. Right: Same figure but for $R$ vs fraction at
             $R$.}
    \label{galactocen_cartesian_radius_vs_fraction}
  \end{figure}

  In \autoref{galactocen_cartesian_radius_vs_fraction}, we show the
  fraction of stars included in the main group of the substructures
  as a function of $|Z|$ and $R$
  \citep[we note that ][shows very similar figures for their more
                        distant red giant sample, see their Figure
                        19]%
  {Naidu_2020}.
  In \autoref{galactocen_cartesian_radius_vs_fraction}, we see that
  Gaia-Enceladus dominates our samples at
  $2\lesssim|Z|\,\text{kpc}^{-1}\lesssim 9$
  and $8\lesssim R\,\text{kpc}^{-1}\lesssim 15$, but that its
  fraction decreases as $|Z|$ or $R$ increase. However, this
  fraction is directly connected to the limitations of extending the
  solar neighbourhood analysis by \citet{Dodd_2023} to these larger
  distances and Gaia-Enceladus being defined from the main body from
  the clustering but does not capture the full Gaia-Enceladus
  extent, for instance to higher energies. With our
  3.2--22.8\,kiloparsecs sample of BHB candidates we sample a wider
  range of families of orbits for substructures compared to just
  those that make it into the limited 2.5\,kiloparsecs radius from
  the Sun studied by
  \citet{Dodd_2023}. This can immediately be appreciated from the
  lower panels of \autoref{results}, where we see that the full BHB
  sample probes a much more extended range of $L _ \perp$ compared
  to the sample by \citet{Dodd_2023}. Massive merger events, such as
  Gaia-Enceladus, are expected to leave debris with a wide range of
  angular momentum and energy, as is shown by simulations of such
  events
  \cite[e.\,g.][]{%
    koppelman_et_al2020, amarante_et_al2022,
    belokurov_et_al2023,sergey_khoperskov_et_al2023_677a90k,
    guillaume_f_thomas_et_al2025}.
  Based on the findings by \citet{balbinot_and_helmi2021}, the
  majority of low $\left|L _ Z\right|$ and high-energy unassigned
  BHB candidates could in fact be part of Gaia-Enceladus that may be
  also linked to some of the retrograde low $\left|L _ Z\right|$ and
  high-$L _ \perp$ unassigned BHB candidates. To a lesser extent some
  of the retrograde and prograde high-energy unassigned BHB candidates
  could be also associated with Sequoia \citep{myeong_et_al2019} and the
  Helmi Streams \citep{helmi_et_al1999helmi_streams}, respectively, due
  to their proximity in $\left(E,\ L _ Z,\ L _ \perp\right)$ space to
  these substructures. That could explain part of the asymmetry
  we observe in the $\left(L _ Z,\ L _ \perp\right)$ space of
  unassigned BHB candidates.

  The most important reason we did not apply the clustering algorithm
  to our sample of BHB stars is the sample bias and incompleteness that
  could cause clusters or artificial overdensities. This would then be
  combined with larger uncertainties, compared to the sample used in
  \citet{Dodd_2023}, which would mean we can define clusters with much
  less certainty. We furthermore note that \autoref{results} also shows
  BHB candidates that are on approximately circular orbits. Their
  current distances would not place most of these targets within
  the thin or thick disc scale heights though instead, they have
  $|Z|\geq 0.8$\,kiloparsec suggesting that perhaps these represent
  mis-classified more nearby (blue straggler) stars. However, the
  targets at 0.8\,kiloparsec above the Galactic plane could be still
  part of the thick disc. Follow-up spectroscopy would be needed to
  confirm their true nature.

  \subsection{Tracing the Sagittarius stream with the deep sample}%
  \label{sagittarius_stream}

\begin{figure}
    \includegraphics{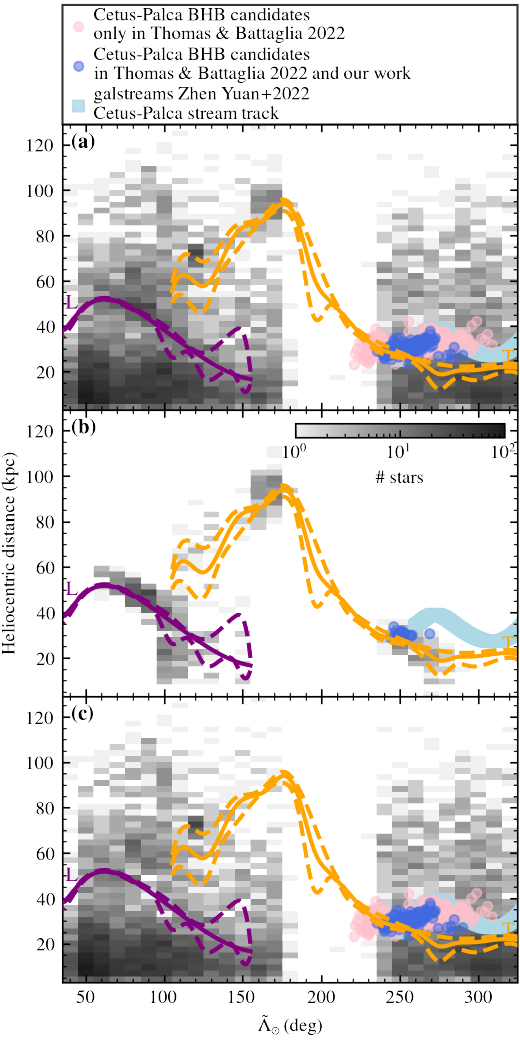}
    \caption{(a) BHB candidates with $P_\text{BHB}\geq 0.5$ and
             $E\left(B-V\right) < 0.1$ in the deep dataset in
             Sagittarius stream coordinate, $\tilde{\Lambda} _ \odot$,
             vs heliocentric distance. The solid and dashed purple and
             orange lines show the leading and trailing arms of the
             stream defined by \citeauthor{hernitschek_et_al2017}
             (\citeyear{hernitschek_et_al2017}, see their tables 4 and
             5), and twice the heliocentric distance uncertainties in 
             each direction, respectively. (b) Stars selected inside
             the defined Sagittarius stream region, based on
             $\left|\tilde{B}_\odot\right| < 15\degr$ and the distance
             mask around the arms. (c) All remaining halo stars.}
    \label{sagittarius_stream_results}
  \end{figure}

  As the Sagittarius stream represents the most dominant coherent
  structure in the outer halo
  \citep[e.g.][]{ibata_et_al2001, majewski_et_al2003, Naidu_2020,
                  johnston_et_al2005, manuel_bayer_et_al2025},
  with a distinct velocity distribution 
  (e.g. \cite{ibata_et_al1994, belokurov_et_al2014,
               ibata_et_al2020, p_ramos_et_al2022}
  Garc{\'i}a~Jim{\'e}nez et al. in prep.),
  we use our deep BHB candidate sample to select (and where
  necessary exclude) members of this stream.  To do so, we converted
  the coordinates into Sagittarius stream longitude
  $\tilde{\Lambda} _ \odot$ and Sagittarius stream latitude
  $\tilde{B} _ \odot$, using the method described in
  \citet[][see their A.1 for the complete coordinate transformation]%
  {belokurov_et_al2014}.
  We subsequently select stars with
  $\left|\tilde{B}_\odot\right| < 15\degr$ and used the Sagittarius
  stream leading and trailing arms as defined by
  \citet{hernitschek_et_al2017} within a window of two standard
  deviations across heliocentric distance to select BHB candidates
  that are within their individual three standard deviations of this
  window. The process is illustrated in
  \autoref{sagittarius_stream_results}.

  Out of 8,043 BHB candidates with $P_\text{BHB}\geq 0.5$ and
  $E\left(B-V\right) < 0.1$, we are left with 7,371 stars outside the
  Sagittarius stream that will be used in all further sections to
  investigate remaining halo substructures without being dominated by
  this prominent structure. We note here that the apparent overdensity
  of BHB candidates in panel (c) at
  $75\degr \leq \tilde{\Lambda} _ \odot \leq 110\degr$ and heliocentric
  distances 47.75--80.00\,kpc can be partly explained by the groupings
  of BHB candidates in the Bo{\"o}tes (I) dwarf galaxy
  \citep{belokurov_et_al2006booetes}.

  \subsection{Characterising the Pisces Overdensity with the deep
              sample}%
  \label{pisces_overdensity}
  
   The Pisces Overdensity at
   $-15\degr \lesssim \text{Right Ascention} \lesssim 5\degr$,
   |Declination| $\leq 1.27$\degr, and distances of 62.5--100.0\,kpc
   was first found by \citet{sesar_et_al2007} on the basis of RR Lyrae,
   who called it feature J (see their figures 12-13). It was later
   coined as the Pisces Overdensity by
   \citet{l_l_watkins_et_al2009} and \citet{kollmeier_et_al2009b}, who
   found similar results with their RR Lyrae samples.
   \citet{belokurov_et_al2019} used BHB stars in addition to RR Lyrae
   in a broader right ascension and distance range from -30\degr\ to
   10\degr, and 40 to 110\,kpc, respectively, and proposed that the
   Pisces Overdensity is linked to the LMC wake
   \citep[see also e.g.][]{amarante_et_al2024}. In their
   exploration, \citet{vedant_chandra_et_al2023a} additionally related
   the retrograde red giant branch stars they selected in the Pisces
   Overdensity to distant apocentric shells of the Gaia-Enceladus
   merger, suggesting that the overdensity might in fact have multiple
   origins. 
  
  To select the Pisces overdensity, we followed
  \citet{amarante_et_al2024} and used the coordinates
  $64\degr\leq l\leq 84\degr$ and $-57\degr\leq b\leq -37\degr$ (their
  table 1). We first used all candidate A-type stars from the
  Pristine-SDSS match to map the observed sky area (and gaps in the
  footprint) by creating a two-dimensional grid in Galactic coordinates
  ($l$, $b$) with a fixed angular size of $l = b = 0.9\degr$ and mark
  each cell with at least one A-type star as `occupied'. We
  subsequently calculated the area of the effective footprint by adding
  the occupied cell areas using
  ${\text{cell area} = l\cdot b\cdot \cos\left(b_\text{centre}\right)}$,
  where $b_\text{centre}$ is the latitude at the cell’s centre.

\begin{figure}
    \includegraphics{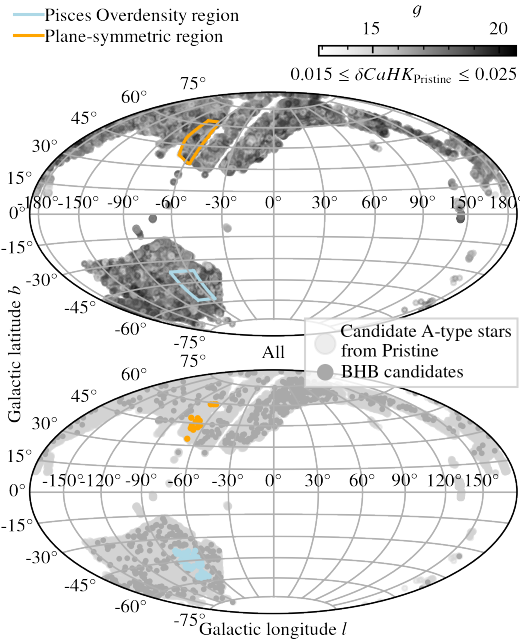}
    \caption{Top panel: Definition of Pisces and symmetric regions in
             Galactic coordinates. The figure shows the depth of the
             selection of BHB candidates. The $g$ colour code represents
             stars with \textit{CaHK} photometric uncertainties
             $0.015\leq\delta CaHK_\text{Pris}\leq 0.025$ in the
             background. The Pisces region is in (light) blue, and the
             symmetric region with respect to the Galactic plane is in
             orange. Bottom panel: Candidate BHB stars (dark grey)
             within 60–100\,kpc. (Light) Blue stars are in the Pisces
             area, and orange stars are in the symmetric area with
             respect to the Galactic plane.}
    \label{pisces_overdensity_region}
  \end{figure}

  Using this area, we selected a region of matching area on the opposite
  side of the Galactic plane with the same longitude range but flipped
  latitudes ($b\to -b$). We then gradually increased the latitude range
  until the total area of `occupied' cells in the Galactic-plane-%
  symmetric region matched that of the Pisces area. In the top
  panel of \autoref{pisces_overdensity_region}, we mark both regions in
  Galactic coordinates with all candidate A stars from the Pristine-SDSS
  match. The colour-coding in this figure corresponds to the depth in
  that part of the sky in $CaHK$ and shows that both regions are
  surveyed to very similar depth. The bottom panel in turn shows only
  the candidate BHB stars between 60 and 100\,kpc.
  
\begin{figure}
    \includegraphics{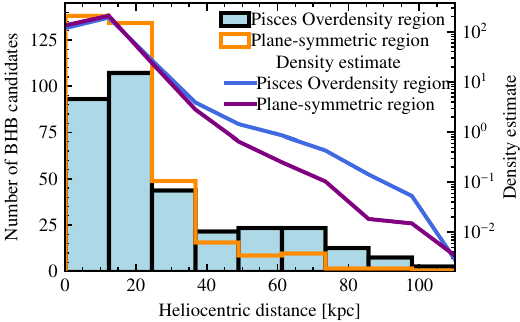}
    \caption{Distribution of heliocentric distances for candidate BHB
             stars in the Pisces and symmetric regions with respect to
             the Galactic plane is shown. The raw star counts are
             represented by bars, while the density estimates are
             represented by solid lines. The density estimates are based
             on a \citet{epanechnikov1969} kernel with a bandwidth of
             16\,kpc.}
    \label{pisces_overdensity_histogram}
  \end{figure}

  Figure \ref{pisces_overdensity_histogram} shows the resulting
  histograms of BHB candidates in heliocentric
  distances from zero to 110\,kpc for each region. The bin width of
  12\,kpc is chosen to be somewhat larger than the typical uncertainty
  in distance. To see trends clearly, we used a density estimation with
  an \citet{epanechnikov1969} kernel weighted by the inverse variance of
  the heliocentric distance. Between 40 and 110\,kpc, there are
  significantly more candidate BHB stars in the Pisces region than in
  the symmetric region with respect to the Galactic plane
  ($50\!\pm\! 8$ vs $14\!\pm\! 4$). At $0.45\!\pm\! 0.07$ stars per
  square degree, the Pisces region is about 2.9 times denser in
  candidate BHB stars for this distance range than the symmetric region
  with respect to the Galactic plane ($0.15\!\pm\! 0.05$ stars per
  square degree) that was observed with a similar depth as the Pisces
  region in Pristine. To test statistical significance of the
  overdensity in Pisces, we computed how many standard deviations the
  two densities differ, which amounts to 3.6, indicating a significance
  above the 3$\sigma$.
  
  We note that although the symmetric region with respect to the
  Galactic plane is less dense in these outer regions, it is denser at
  closer distances, as can be seen from
  \autoref{pisces_overdensity_histogram}, due to the overlap with
  the Hercules-Aquila Cloud, which dominates at distances of order
  20\,kpc
  \citep[e.\,g.][]{%
    belokurov_et_al2007hercules_aquila_cloud, l_l_watkins_et_al2009,
    jelte_de_jong_et_al2010, vedant_chandra_et_al2023a,
    amarante_et_al2024}.
  Therefore, for the significance test at the distances of the Pisces
  Overdensity we still compare against a background in the
  Galactic-plane-symmetric regions devoid of the Hercules-Aquila
  Cloud.
        
  Further kinematic exploration of our sample of BHB candidates through 
  follow-up spectroscopy (as currently only 16 of the 41 stars have
  radial velocities) and improved proper motions with Gaia DR4 could
  help to disentangle the dominant formation history of this
  overdensity by investigating whether their orbits are retrograde and
  would be potentially consistent with distant Gaia-Enceladus debris
  rather than part of the LMC wake. 
      
  \subsection{Comparison of the bulk halo motion between hemispheres}%
  \label{bulk_halo_motion}
  
  Among the many substructures in the outer halo, simulations predict
  both a Galactic north-south line-of-sight velocity asymmetry and
  stellar overdensities in the outer halo as response to the infall of
  the Large Magellanic Cloud (LMC)
  \cite[e.g.][]{%
    garavito_camargo_et_al2019, michael_s_petersen_and_penarrubia2020, 
    garavito_camargo_et_al2021}.
  The asymmetry is a result of the Milky Way reflex motion where the
  Galactic disc and inner halo move towards the Galactic south. This was
  already observed by
  \citet{michael_s_petersen_and_penarrubia2021} - using a combination
  of K-type giants, BHB stars, and satellite galaxies - and has since
  been extended in many subsequent studies, including
  \citet{bystroem_et_al2025}, who used a large dataset of DESI BHB
  stars.
  
  In this work, we used our deep sample to look into this bulk motion
  as a function of distance. We implemented a GSR line-of-sight velocity
  uncertainty cut of 20\,km\,s$^{-1}$ to eliminate any stars with
  uncertain velocities while maintaining a substantial sample size.
  This reduction decreases the sample from 3,952 stars to 3,837, with
  most of the excluded stars located at the largest distances.

  For our analysis, we divided our sample of BHB candidates with
  distances $>$ 10 kpc into five bins with bin boundaries 10, 15, 20,
  30, 50, and 107\, kpc to get a more granular look at the data where
  our sample is densest but retaining good statistics in the sparser
  outer regions. We then computed the mean velocities for the set of
  stars in each bin, $\langle v \rangle$, and standard deviation,
  $\sigma$, by maximising the log-likelihood following the method
  applied by \citet{bystroem_et_al2025}:
  \[-\frac{1}{2}\sum_{i = 1}^N%
  \left(%
  \ln\left(2\pi\left(\sigma^2+\delta v_{i,\text{h}}^2\right)\right)%
  + \frac{\left(v_{i,\text{h}}-\langle v\rangle\right)^2}%
  {\sigma^2+\delta v_{i,\text{h}}^2}\right).\]
  This accounts for the individual uncertainties,
  $\delta v_{i,\text{h}}$, on the velocities by using
  $\sigma^2 + \delta v_{i,\text{h}}^2$ instead of just $\sigma^2$. To
  perform the maximisation, we used the negative log-likelihood and
  SciPy's \citep{scipy} \texttt{optimize.minimize} with the
  limited-memory Broyden-Fletcher-Goldfarb-Shanno method with simple
  bounds
  \citep{richard_h_byrd_et_al1995, c_zhu_et_al1997algorithm, 
         j_l_morales_and_nocedal2011remark}.
  The uncertainties on $\langle v \rangle$ and $\sigma$ were obtained by
  taking the square root of the diagonals of the covariance matrix,
  which we computed from the Hessian matrix.
  
\begin{figure}
    \includegraphics{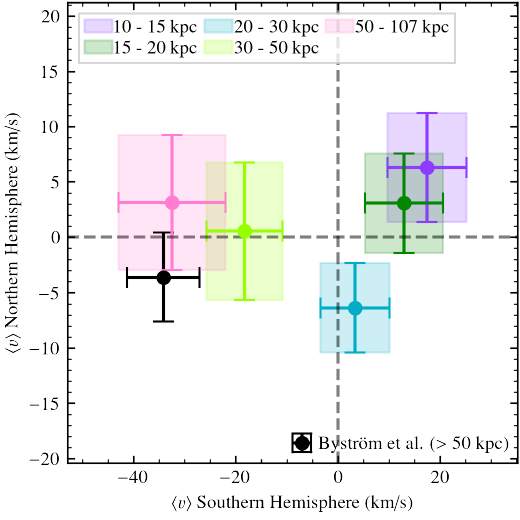}
    \caption{Average Galactocentric line-of-sight velocity in the
             Galactic northern versus the southern hemispheres for each
             bin between 10 and 107\,kpc, excluding the Sagittarius
             stream. For comparison, the results beyond 50\,kpc from
             \citet{bystroem_et_al2025} are shown as well.}
    \label{north_vs_south}
  \end{figure}

  In \autoref{north_vs_south} the resulting means in each bin are
  shown, with the southern hemisphere on the x axis and the northern
  hemisphere on the y axis. A clear trend emerges when examining how
  the southern hemisphere velocity changes with Galactic radius. The
  four innermost bins, spanning from 10 to 50\,kpc, are clustered near
  the origin, with a slight inclination towards a positive southern
  hemisphere line-of-sight velocity. As we extend further into the
  Galaxy, we observe a shift in the southern hemisphere velocity,
  becoming increasingly negative as we move farther out. Our result for
  the most distant bin of $-$32 $\pm$ 11\,km\,s$^{-1}$ agrees with the
  findings of \citet{bystroem_et_al2025}, as indicated by the black
  symbol in \autoref{north_vs_south}. Interestingly, we also observe a
  shift even closer to the Galactic centre between 30 and 50\,kpc of
  $-$18 $\pm$ 8\,km\,s$^{-1}$.
  
  We note that the value obtained by
  \citet{michael_s_petersen_and_penarrubia2021} of
  $-22 ^ {+8} _ {-8}\,\text{km}\,\text{s}^{-1}$ in their sample of
  southern hemisphere BHBs beyond 40\,kpc agrees well with our findings
  (they also find similar results in their sample of K-type giant
  stars). Similarly, \citet{vedant_chandra_et_al2026}, using a K-giant
  sample from SDSS-V beyond 60\,kpc, find a mean velocity in the
  southern hemisphere of $\sim -30$\,km\,s$^{-1}$. They also perform
  radial binning and observe a similar trend, with the southern
  hemisphere radial velocity becoming more negative at larger radii,
  similar to how we observe the effect starting at 30\,kpc.
  
  \section{Discussion}\label{discussion}
  
  \subsection{The fraction of BHB stars as function of magnitude}%
  \label{magnituderelation}
  
\begin{figure}
    \includegraphics{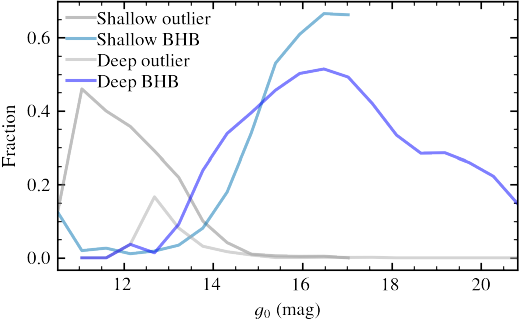}
    \caption{Ratio of BHB (blue) and outlier candidates (grey) as a 
             function of magnitude, as an outcome of the modelling
             (see text for details).}
    \label{fraction_vs_magnitude}
  \end{figure}

  As we do not assume a priori a relation between BHB fraction and
  $g$ magnitude, we can look at this relation as a result of the
  model. The resulting ratios of BHB (blue) and outlier candidates
  (grey) as a function of magnitude are depicted in
  \autoref{fraction_vs_magnitude}. This figure demonstrates that the
  magnitude-dependence of the fractions of the subgroup of BHB and
  outlier candidates cannot be described by a linear function in our
  dataset (we note that this was an assumption in the population
  model of \citealt{fengqing_yu_et_al2024}). The outlier
  fractions peak at brighter magnitudes for both the shallow and
  deep datasets, which is probably related to the presence of young,
  hot main-sequence stars that are more highly represented in the
  shallow dataset footprint. Interestingly, both catalogues show a rise
  in the fraction of BHB stars around $g _ 0$ = 14 which would
  correspond to a heliocentric distance in the range of four to five
  kiloparsecs assuming a BHB star $M_{g} \sim 0.5$--0.7. We note that
  the small differences between the curves and their peak values do not
  necessarily point to inconsistencies between the two catalogues, as
  the footprint covered is somewhat different as well. The deep BHB
  sample then shows a clear drop in the BHB fraction at magnitudes
  larger than $g _ 0$ = 16, which would point to a larger
  fraction of blue stragglers at those magnitudes. Unfortunately, we
  cannot probe these magnitudes with the shallow sample. At faint
  magnitudes, where we would expect possibly more quasar and white
  dwarf contamination into the sample
  \citep[e.g.][]{%
    Yanny_2000, Deason2011, hugh_c_harris_et_al2003astronomical_journal,
    deason_et_al2014, gentile_fusillo_et_al2015a, montegriffo_et_al2023,
    o_vincent_et_al2024}
  the outlier fractions are practically zero.
      
  \subsection{Method comparison with previous work}

  The method developed in this work is based on the classification
  approach used in \citet{fengqing_yu_et_al2024}, but using the
  colour-colour space from \citet{Starkenburg2019}. However, there
  are significant differences with both these approaches. A key
  distinction between our approach and that of
  \citet{Starkenburg2019} lies in the definition of the BHB and blue
  straggler sequences. In our method, these sequences are
  characterised by natural dispersions that we also derive from the
  data, rather than relying on pre-set values (such as
  for instance those from \citealt{Deason2011} that are used in
  \citealt{Starkenburg2019}). Notably, the best estimate of
  $\sigma _ \text{BHB}$ obtained in our study (0.024) seems quite
  different from the 0.04 value assumed in \citet{Starkenburg2019}.
  However, upon reviewing the distribution of
  the sampled values of $\sigma _ \text{BHB}$ in
  \autoref{corner_plot_training}, the difference fails to reach
  statistical significance. On the other hand, the inferred
  $\sigma _ \text{straggler}$ in our study (0.044) aligns almost
  perfectly with the value adopted by
  \citet[][which is 0.045]{Starkenburg2019}.

  Our model also accommodates for the possibility that a star belongs to
  neither the BHB nor blue straggler populations, which is not
  feasible in a normalised two-class framework as employed in
  \citet{Deason2011} and \citet{Starkenburg2019}. This approach
  provides a more reliable estimate of the probability and helps
  avoid erroneous classifications. In terms of performance, both
  methods achieve a comparable completeness of approximately 91\%.
  However, our estimates of the completeness in the lower part of
  \autoref{table_purity_and_completeness} should be
  considered upper limits considering that the test and validation
  samples may themselves contain some degree of impurity and
  incompleteness. For comparison, methods based on variations in the
  SDSS \textit{z} band typically achieve lower completeness levels,
  ranging from around 51.0\% to 96.2\%
  \cite[e.g.][]{Vickers_2012, Fukushima_2018, Thomas2018Atype,
                 fengqing_yu_et_al2024}.
  This highlights the advantage of using the \textit{CaHK} filter.

  In terms of distance reach, \citet{Starkenburg2019} identified
  candidate BHB stars with a probability of at least 0.8 out to
  approximately 80\,kpc. Our sample extends the same probability
  threshold to approximately 100\,kpc. For candidate BHB stars with
  a probability greater than 0.5, their sample reached approximately
  100\,kpc, while ours extended out to approximately 120\,kpc. This
  demonstrates that our method can identify BHB stars with higher
  confidence to fainter magnitudes.
    
  For our shallow sample, it is instructive to compare our results
  to other catalogues of BHB stars using Gaia XP spectra, in
  particular the work of \citet{guozhen_hu_et_al2025} and
  \citet{wenbo_wu_et_al2025} already mentioned in
  \autoref{pristine_gaia_uband_section}. If we take the catalogues
  by \citet{guozhen_hu_et_al2025} and \citet{wenbo_wu_et_al2025} at
  face value without any quality cuts, we find that, as anticipated,
  our method adds BHB candidates towards the faint part of the set
  of \textit{Gaia} DR3 sources for which XP spectra are available
  because we use the deeper SDSS \textit{u} data (see
  \autoref{other_catalogues} for more details). With the added
  caveat that the quality of the photometry and possibly
  probabilistic photometric classification will decrease with
  decreasing $b$, we would like to note in terms of sky coverage
  that we also add BHB candidates for $|b| < 8\degr$.

  \section{Conclusion and future outlook}\label{conclusion}
  
  In this paper, we extended the hierarchical Bayesian mixture model
  of \citet{fengqing_yu_et_al2024} for separating candidate BHB from
  candidate blue straggler stars, using photometric data from the
  Pristine and SDSS surveys. We assigned probabilistic classifications
  in the $\left((u-CaHK) _ 0,\ (g-r) _ 0\right)$ colour-colour space
  following \citet{Starkenburg2019}, which clearly separates the two
  stellar populations. This approach enabled us to identify
  candidate BHB stars at larger distances and with high confidence,
  even in the presence of significant photometric uncertainties.
  
  We applied this method to create two distinct samples: a shallow
  dataset using the Pristine-\textit{Gaia} synthetic catalogue
  \textit{CaHK} and a deep dataset using Pristine \textit{CaHK}. For
  both datasets, each star received a final classification
  probability of being BHB, blue straggler, or outlier. Selecting
  stars with $P_\text{BHB}\geq0.5$, we built a large BHB candidate
  catalogue that includes more than 5,100 stars in the shallow
  dataset and 8,824 stars in the deep dataset, some extending beyond
  110\,kpc. More than 9,200 of these stars have high classification
  probabilities $\left(P_\text{BHB}\geq 0.8\right)$ and the upper
  limits of the purity and completeness compared to known
  spectroscopic samples reach 100\% and 99\%, and 93\% and 94\%, for the
  shallow/deep datasets, respectively. In addition, we matched our
  $P_\text{BHB}\geq 0.5$ BHB samples with the DESI, LAMOST,
  \textit{Gaia} RVS, GALAH, and SEGUE spectroscopic surveys, obtaining
  line-of-sight velocity measurements for over 6,100 stars.
  
  Using the shallow dataset, we traced known stellar structures in
  $\left(L_Z,\ E,\ L_\perp\right)$ dynamical-quantity space,
  including Gaia-Enceladus, the Helmi Streams, and Thamnos, and found
  that out of 2,769 BHB candidates, 437 are assigned to
  substructures, with Gaia-Enceladus dominating the sample at
  $2\lesssim\left|Z\right|\,\text{kpc}^{-1}\lesssim 9$ and
  $8\lesssim R\,\text{kpc}^{-1}\lesssim 15$. The full BHB sample
  probes a wider range of orbits compared to previous studies and we
  only assign BHB candidates to the main overdensities in
  \citet{Dodd_2023} without their full extent, suggesting that many
  unassigned candidates could be part of Gaia-Enceladus or other
  substructures. Our analysis of the deep dataset maps large
  substructures in the outer halo, most notably the Sagittarius stream
  and the Pisces Overdensity. The added line-of-sight velocities in the
  deep dataset allowed us to study the kinematics of the halo and
  compare the velocity distributions in the northern and southern
  Galactic hemispheres. The results showed clear and significant
  velocity asymmetries between the two hemispheres, corroborating
  literature results by \citet{michael_s_petersen_and_penarrubia2021},
  \citet{vedant_chandra_et_al2026}, and \citet{bystroem_et_al2025}.
    
  For the shallow dataset, based on currently available XP
  spectra for roughly 200 million stars, we are restricted to
  17.65\,mag in $G$, but we foresee large potential in future Gaia
  releases, where the XP spectra of 2 billion sources (down to
  20.5\,mag in $G$) will become available. When such data become
  available, we suggest a future investigation into the use of
  alternative colour-colour spaces that do not depend on any
  variation in the SDSS \textit{u} filter, enabling an all-sky
  view. A first attempt in this direction is presented in
  \autoref{section_colour_colour_diagrams}. While here we propose
  two candidate colour-colour diagrams using J-PAS filters, they do
  not perform better than the method applied in this work. More
  general searches with modern techniques such as random forest or
  more flexible custom filter combinations with \texttt{GaiaXPy},
  might further this endeavour. We like to note, however, that
  future \textit{Gaia} data will still be limited at the bluest end,
  before the Balmer jump, limiting any method that uses these
  wavelength regions to the higher-S/N stars. 
  
  Additionally, improved Gaia DR4 proper motions combined with more
  radial velocities from surveys such as the DESI, William Herschel
  Telescope Enhanced Area Velocity Explorer
  \citep{dalton_et_al2012, dalton_et_al2016, dalton2016,
         shoko_jin_et_al2024}
  and 4-metre Multi-Object Spectroscopic Telescope
  \citep{r_s_de_jong_et_al2019} will enable full orbital modelling
  of the BHB candidates assigned to stellar streams and
  overdensities and increase our understanding of their origins.
  Overall, this research showcases the remarkable power of photometric
  data in identifying potential BHB stars, even in the face of
  substantial uncertainties and at distances exceeding 100\,kpc.
  
  \section*{Data availability}
  
  The catalogue described in \autoref{dataset_description} is only
  available in electronic form at the CDS via anonymous ftp to
  \url{cdsarc.u-strasbg.fr} (130.79.128.5) or via
  \url{http://cdsweb.u-strasbg.fr/cgi-bin/qcat?J/A+A/}.
  
\begin{acknowledgements}
  The authors like to thank the anonymous referee for a careful and
  constructive report that has certainly helped to improve the
  manuscript. The authors would like to thank Martin~Montelius,
  Thomas~M.~Callingham, and Samuel~Rusterucci for helpful comments that
  helped to improve this study. We want to express our gratitude to
  Eduardo~Balbinot for providing us the photometry from separate
  synthetic photometric bands (supplied by the GSPC or determined
  using \texttt{GaiaXPy}). We also would like to thank Eduardo~%
  Balbinot and Tom{\'a}s Ruiz-Lara for arranging the \textit{Gaia}
  (E)DR3 catalogues and pertinent survey matches to be accessible to the
  `Galactic' group members. ES and MB acknowledge funding through
  VIDI grant `Pushing Galactic Archaeology to its limits' (with
  project number VI.Vidi.193.093) which is funded by the Dutch
  Research Council (NWO). This research has been partially funded
  from a Spinoza award by NWO (SPI 78-411). ES acknowledges funding
  from the European Research Council (ERC, EARLYMW, Grant number
  101170507). Views and opinions expressed are those of the
  author(s) only and do not necessarily reflect those of the
  European Union or the European Research Council. Neither the
  European Union nor the granting authority can be held responsible
  for them. GFT acknowledges support from the Agencia Estatal de
  Investigaci\'on del Ministerio de Ciencia en Innovaci\'on (AEI-%
  MCIN) under grant number PID2023-150319NB-C21 and the grant
  RYC2024-051016-I funded by MCIN/AEI/10.13039/501100011033 and by
  the European Social Fund Plus. Co-funded by the European Union
  (Widening Participation, ExGal-Twin, GA 101158446). The authors
  acknowledge the use of the Catalog Archive Server Jobs System
  (CasJobs) service at http://casjobs.sdss.org/CasJobs, developed by
  the JHU/SDSS team. This work uses SDSS and SDSS SEGUE data.
  Funding for the Sloan Digital Sky Survey IV has been provided by
  the Alfred P. Sloan Foundation, the U.S. Department of Energy
  Office of Science, and the Participating Institutions
  SDSS-IV acknowledges support and resources from the Center for
  High Performance Computing  at the University of Utah. The SDSS
  website is www.sdss4.org.
  SDSS-IV is managed by the Astrophysical Research Consortium for
  the Participating Institutions of the SDSS Collaboration including
  the Brazilian Participation Group, the Carnegie Institution for
  Science, Carnegie Mellon University, Center for Astrophysics |
  Harvard \& Smithsonian, the Chilean Participation Group, the
  French Participation Group, Instituto de Astrof\'isica de
  Canarias, The Johns Hopkins University, Kavli Institute for the
  Physics and Mathematics of the Universe (IPMU) / University of
  Tokyo, the Korean Participation Group, Lawrence Berkeley National
  Laboratory, Leibniz Institut f\"ur Astrophysik Potsdam (AIP),
  Max-Planck-Institut f\"ur Astronomie (MPIA Heidelberg),
  Max-Planck-Institut f\"ur Astrophysik (MPA Garching),
  Max-Planck-Institut f\"ur Extraterrestrische Physik (MPE),
  National Astronomical Observatories of China, New Mexico State
  University, New York University, University of Notre Dame,
  Observat\'ario Nacional / MCTI, The Ohio State University,
  Pennsylvania State University, Shanghai Astronomical Observatory,
  United Kingdom Participation Group, Universidad Nacional
  Aut\'onoma de M\'exico, University of Arizona, University of
  colourado Boulder, University of Oxford, University of Portsmouth,
  University of Utah, University of Virginia, University of
  Washington, University of Wisconsin, Vanderbilt University, and
  Yale University. 
  The Pan-STARRS1 Surveys (PS1) and the PS1 public science archive have
  been made possible through contributions by the Institute for
  Astronomy, the University of Hawaii, the Pan-STARRS Project Office,
  the Max-Planck Society and its participating institutes, the Max
  Planck Institute for Astronomy, Heidelberg and the Max Planck
  Institute for Extraterrestrial Physics, Garching, The Johns Hopkins
  University, Durham University, the University of Edinburgh, the
  Queen's University Belfast, the Harvard-Smithsonian Center for
  Astrophysics, the Las Cumbres Observatory Global Telescope Network
  Incorporated, the National Central University of Taiwan, the Space
  Telescope Science Institute, the National Aeronautics and Space
  Administration under Grant No. NNX08AR22G issued through the
  Planetary Science Division of the NASA Science Mission Directorate,
  the National Science Foundation Grant No. AST-1238877, the University
  of Maryland, Eotvos Lorand University (ELTE), the Los Alamos National
  Laboratory, and the Gordon and Betty Moore Foundation. Based on
  observations obtained at the Canada-France-Hawai'i Telescope (CFHT)
  which is operated by the National Research Council of Canada, the
  Institut National des Sciences de l'Univers of the Centre National de
  la Recherche Scientifique of France, and the University of Hawai'i.
  CFHT is located on Maunakea on Hawai'i Island, a mountain of
  considerable cultural, natural, and ecological significance. Maunakea
  is a sacred site to Native Hawaiians, also known as Kānaka 'Ōiwi.
  Quality observations are made possible by relentless effort of the
  entire staff at Canada-France-Hawai'i Telescope. Based on
  observations obtained with MegaPrime/MegaCam, a joint project of CFHT
  and CEA/DAPNIA. This work has made use of data from the European
  Space Agency (ESA) mission {\it Gaia}
  (\url{https://www.cosmos.esa.int/gaia}), processed by the {\it Gaia}
  Data Processing and Analysis Consortium (DPAC,
  \url{https://www. cosmos.esa.int/web/gaia/dpac/consortium}). Funding
  for the DPAC has been provided by national institutions, in
  particular, the institutions participating in the {\it Gaia}
  Multilateral Agreement. 
  This work made use of the Fourth Data Release
  of the GALAH Survey \citep{buder_et_al2021survey_data_release3}. The
  GALAH Survey is based on data acquired through the Australian
  Astronomical Observatory, under programmes: A/2013B/13 (The GALAH
  pilot survey); A/2014A/25, A/2015A/19, A2017A/18 (The GALAH survey
  phase 1); A2018A/18 (Open clusters with HERMES); A2019A/1
  (Hierarchical star formation in Ori OB1); A2019A/15, A/2020B/23,
  R/2022B/5, R/2023A/4, R2023B/5 (The GALAH survey phase 2);
  A/2015B/19, A/2016A/22, A/2016B/10, A/2017B/16, A/2018B/15 (The
  HERMES-TESS programme); A/2015A/3, A/2015B/1, A/2015B/19, A/2016A/22,
  A/2016B/12, A/2017A/14, A/2020B/14 (The HERMES K2-follow-up
  programme); R/2022B/02 an A/2023A/09 (Combining asteroseismology and
  spectroscopy in K2); A/2023A/8 (Resolving the chemical fingerprints
  of Milky Way mergers); and A/2023B/4 (s-process variations in
  southern globular clusters).
  We acknowledge the traditional owners of the land on which the AAT
  stands, the Gamilaraay people, and pay our respects to elders past
  and present. This paper includes data that has been provided by AAO
  Data Central (datacentral.org.au). 
  This research used data obtained with the Dark Energy Spectroscopic
  Instrument (DESI). DESI construction and operations is managed by the
  Lawrence Berkeley National Laboratory. This material is based upon
  work supported by the
  \href{https://www.energy.gov/}{U.S. Department of Energy}, Office of
  Science, Office of High-Energy Physics, under Contract No.
  DE–AC02–05CH11231, and by the National Energy Research Scientific
  Computing Center, a DOE Office of Science User Facility under the
  same contract. Additional support for DESI was provided by the
  \href{https://www.nsf.gov/}{U.S. National Science Foundation} (NSF),
  Division of Astronomical Sciences under Contract No. AST-0950945 to
  the NSF’s National Optical-Infrared Astronomy Research Laboratory;
  the
  \href{https://stfc.ukri.org/}{Science and Technology Facilities
                                Council of the United Kingdom};
  the \href{https://www.moore.org/}{Gordon and Betty Moore Foundation};
  the \href{https://www.hsfoundation.org/}{Heising-Simons Foundation};
  the \href{https://www.cea.fr/}{French Alternative Energies and Atomic
                                 Energy Commission}
  (CEA); the
  \href{https://secihti.mx/}{National Council of Humanities, Science
                             and Technology of Mexico}
  (CONAHCYT); the
  \href{http://www.mineco.gob.es/}{Ministry of Science and Innovation
                                   of Spain}
  (MICINN), and by the DESI Member Institutions:
  \url{www.desi.lbl.gov/collaborating-institutions}. The DESI
  collaboration is honored to be permitted to conduct scientific
  research on I’oligam Du’ag (Kitt Peak), a mountain with particular
  significance to the
  \href{https://www.tonation-nsn.gov/}{Tohono O’odham Nation}. Any
  opinions, findings, and conclusions or recommendations expressed in
  this material are those of the author(s) and do not necessarily
  reflect the views of the U.S. National Science Foundation, the U.S.
  Department of Energy, or any of the listed funding agencies.
  Guoshoujing Telescope (the Large Sky Area Multi-Object Fiber
  Spectroscopic Telescope LAMOST) is a National Major Scientific Project
  built by the Chinese Academy of Sciences. Funding for the project has
  been provided by the National Development and Reform Commission.
  LAMOST is operated and managed by the National Astronomical
  Observatories, Chinese Academy of Sciences. This research has made use
  of the VizieR catalogue access tool, CDS, Strasbourg, France (DOI:
  10.26093/cds/vizier). The production of this work made use of the
  condor distributed computing software. This research has made use of
  the SVO Filter Profile Service ``Carlos Rodrigo'', funded by
  MCIN/AEI/10.13039/501100011033/ through grant PID2023-146210NB-I00.
  This work made use of the software Action-based Galaxy Modelling
  Architecture \citep{vasiliev2018, vasiliev2019},
  \texttt{astroML} \citep{vanderplas_et_al2012}, \texttt{Astropy}
  \citep{astropy2013, astropy2018, astropy2022}, \texttt{astroquery}
  \citep{ginsburg_et_al2019astroquery}, \texttt{colorcet}
  \cite[colorcet.com,][]{kovesi2015arxiv1509_03700}, \texttt{dustmaps} 
  \citep{green2018}, \texttt{dynesty}
  \citep{speagle2020}, \texttt{GaiaXPy}
  (gaia-dpci.github.io/GaiaXPy-website), \texttt{Gala} \citep{gala},
  \texttt{galstreams} \citep{mateu_et_al2018}, JupyterLab
  \citep{kluyver_et_al2016}, \texttt{Kapteyn}
  \citep{terlouw_and_vogelaar2016}, \texttt{Matplotlib},
  \citep{hunter2007}, \texttt{NumPy} \citep{numpy}, \texttt{pandas}
  \citep{mckinney-proceeding-scipy-2010, reback2020pandas},
  \texttt{Polars}, \texttt{pystellibs}
  \url{mfouesneau.github.io/pystellibs/}, \texttt{Python} (python.org),
  \texttt{Scikit-learn} \citep{scikit_learn}, \texttt{seaborn}
  \citep{waskom2021}, Starlink Tables Infrastructure Library Tool Set
  \citep{m_b_taylor2006}, \texttt{schwimmbad}
  \citep{price_whelan_and_foreman_mackey2017}, Tool for Operations on
  catalogues and Tables \citep{m_b_taylor2005}, \texttt{tqdm}
  \citep{casper_da_costa_luis2019}, and \texttt{Vaex}
  \citep{breddels_and_veljanoski2018}. We would like to
  acknowledge the use of DeepL Write, an artificial intelligence
  writing assistant developed by DeepL, and Apple Intelligence which was
  employed to enhance the readability and style of this article and
  for textual editing for this article. We used the Mistral Small 3/4
  chat model, OpenAI's chat model 4/5 based on the generative pre-%
  trained transformer, and Claude Haiku 4.5 chat model through the
  DuckDuckGo AI chat assistant to edit the draft and sought guidance for
  developing Python code for the data analysis and plotting. Author
  contributions: All authors helped in finalising the paper draft.
  Conceptualisation: M. Bayer, E.S., A.Z., F.M., M. Bohak, N.F.M., E.D.,
  H.C.W., G.F.T., A.H. Methodology: M. Bayer, E.S., A.Z., F.M., M.
  Bohak, N.F.M., E.D., H.C.W., G.F.T., A.H. Software: M. Bayer, E.S.,
  A.Z., F.M., M. Bohak, E.D., H.C.W. Validation: M. Bayer, E.S., A.Z.,
  F.M. M. Bohak, N.F.M., E.D., H.C.W., A.H. Formal Analysis: M. Bayer,
  E.S., A.Z., F.M., M. Bohak, E.D., H.C.W., A.H. Investigation: M.
  Bayer, E.S., A.Z., F.M., M. Bohak, N.F.M., E.D., H.C.W., G.F.T., A.H.
  Data Curation: M. Bayer, E.S., A.Z., F.M., N.F.M., E.D., H.C.W.,
  G.F.T. Writing - Original Draft: M. Bayer, E.S., A.Z., F.M., M.
  Bohak. Writing - Review \& Editing: M. Bayer, E.S., A.Z., M. Bohak,
  N.F.M., E.D., H.C.W., G.F.T., A.H., J.F.N. Visualisation: M. Bayer,
  E.S., A.Z., F.M., M. Bohak, N.F.M., E.D., H.C.W., G.F.T., A.H., J.F.N.
\end{acknowledgements}

\bibliographystyle{aa}
\bibliography{refs}

\begin{appendix}
  \section{SQL query at Sloan CasJobs to assemble the Sloan
           $u_0 g_0 r_0$ photometric catalogue of candidate A-type
           stars}%
  \label{sql_query}
           
  The following SQL query was used to assemble a SDSS $u_0 g_0 r_0$
  photometric catalogue of 335,247 candidate stars in part of the
  $(u-g)_0$ vs $(g-r)_0$ colour-colour diagram. We followed the
  selection criteria in the query that were also used in
  \citet{Starkenburg2019}, given here below. For the deselection of
  saturated photometry, de-blending problems, interpolation problems,
  suspicious detections, and stars close to the edge of the frame, we
  give the query for the u band, but the same queries have been
  repeated for the g and r band.
           
  \begin{verbatim}
    select
      objid, ra, dec, raErr, decErr, err_u,
      err_g, err_r, dered_u, dered_g,
      dered_r
    into
    mydb.cleanSDSS_DR19ugr_candAtype_stars
    --- primary objects classified as
    --- stars
    from Star 
    where
      --- colour-colour diagram range for
      --- candidate A-type stars
      (dered_u-dered_g > 0.9)
      and (dered_u-dered_g < 1.4)
      and (dered_g-dered_r > -0.25)
      and (dered_g-dered_r < -0.05)
      and (clean = 1)
      --- unsatured ugr photometry
      and ((((flags_u & 0x40000) = 0)
       --- no de-blending problems
           and (
             (((flags_u & 0x20) = 0)
              and ((flags_u & 0x80000)
                   = 0)
              and (
                ((flags_u 
                  & 0x400000000000)
                 = 0) 
                or (psfmagerr_u <= 0.2))
             )
           )
        --- no interpolation problems
        and (
          (((flags_u
             & 0x800000000000) = 0)
          and (
            (flags_u & 0x10000000000)
            = 0)
          and (
            ((flags_u
              & 0x100000000000)
             = 0)
            or ((flags_u & 0x1000)
                = 0)
           )
          )
        )
        --- no suspicious
        --- detections
        and (
          (((flags_u
             & 0x10000000)
            != 0)
           and (
             (flags_u
              & 0x40000)
             = 0)
           and (
             (flags_u
              & 0x80)
             = 0)
          ) 
         )
        --- not close to
        --- edge of a
        --- frame
        and (
          ((flags_u
            & 0x4)
           = 0)
        )
      )
  \end{verbatim}
  
  \section{\textit{Gaia}-based amplitude of variability for blue
           horizontal branch stars}%
  \label{variability_appendix_section}
           
\begin{figure}
    \includegraphics{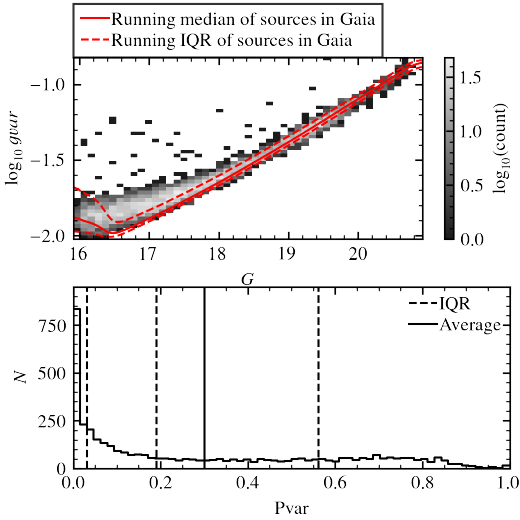}
    \caption{\textit{Gaia}-based variability in \textit{G} shown for
             the BHB stars in the value-added catalogue in the DESI
             \citep{desi_edr, desi_dr1} MWS
             \citep{andrew_p_cooper_et_al2023, koposov_et_al2024,
                    koposov_et_al2026}
             DR1 compiled in \citet{bystroem_et_al2025}. The top panel
             shows the variability distribution in the magnitude range
             of the BHB stars. We give the summary statistics of
             P$_\text{var}$ of the BHB stars that result from the model 
             inferred in \pristine\ (\citeyear{PristineDR1}) in the
             bottom panel in the form of a histogram and overlaid
             quantities. The quantities we include are the
             interquartile range (plus median; dashed lines) and the
             average (solid line).}
    \label{variability_distr_results}
  \end{figure}

  The \textit{Gaia}-based variability metric g$_\text{var}$ defined in
  \citeauthor{belokurov_et_al2017}
  $\left(\text{\citeyear{belokurov_et_al2017}, the authors use }%
  Amp = \log_{10}\left(\text{g}_\text{var}\right)\right)$ and used in
  \pristine\ (\citeyear{PristineDR1}) to remove variable sources from
  (synthetic) \textit{CaHK} photometric catalogues of FGK stars is
  higher for BHB stars at the bright end ($G \lesssim 17.5$) in the
  value-added catalogue \cite[compiled in][]{bystroem_et_al2025} in the
  DESI MWS DR1 than we would expect from the median spread of the
  distribution of the sources in \textit{Gaia} DR3 in the same
  range of g$_\text{var}$ and $G$ as the BHB stars (see the top panel of
  \autoref{variability_distr_results}). This variability distribution
  also leads to a broad range of P$_\text{var}$, which means that
  the adopted variability cut of P$_\text{var} < 0.3$ in \pristine\
  (\citeyear{PristineDR1}) would remove many of the BHB stars. The
  bright BHB stars might therefore be not close to the photon noise
  limit of the instrument, but we also observe that the photometric
  uncertainties in the blue end of the spectrum are higher in the
  \textit{Gaia} data set, as reported in \citet{gedr3} and
  \citet{riello_et_al2021}. One explanation for this might be a high
  fraction of binary stars as reported and suggested by
  \citet{belokurov_et_al2020_mnras_496_1922b}
  \cite[see also figure 14 in][]{maiz_apellaniz_et_al2023}. As noted by
  \citet{belokurov_et_al2020_mnras_496_1922b}, the process of red giant
  branch mass loss exerts a pivotal influence on the precise
  positioning of a star along the horizontal branch, thereby dictating
  its subsequent evolutionary trajectory. Furthermore, to date, a
  definitive theoretical explanation for the observed mass loss remains
  elusive. Therefore, given that mass transfer from a stellar
  companion might also effect the degree of mass loss on the red giant
  branch and placement of a star on the horizontal branch depends on
  the red giant branch mass loss, we currently do not have a solid
  understanding on what to expect in terms of binary stars across the
  HB. In sum, as of now we do not understand thoroughly the low-level
  variability of BHB stars measured with g$_\text{var}$.
  
  \section{$CaHK$ synthetic against Pristine for candidate A-type
           stars}%
  \label{appendix_cahk_synthetic_pristine_magnitude_difference}
  
  Figure
  \ref{%
    appendix_figure_cahk_synthetic_pristine_magnitude_difference} shows
  that despite the fact that the Pristine survey $CaHK _ \text{Pris}$
  was calibrated against $CaHK _ \text{synth}$, we observe a systematic
  offset on the order of -0.04 to -0.045 for a sample of A-type stars
  between the two $CaHK$ magnitudes. Additionally, according to the
  Spearman rank correlation coefficient there is a slight degree of a
  positive correlation between the $CaHK$ magnitude difference and
  $(g-r)$. One explanation for this might be that the calibration of
  $CaHK _ \text{Pris}$ against $CaHK _ \text{synth}$ does not
  perform as good for candidate A-type stars as it does for FGK stars
  (see \pristine\ \citeyear{PristineDR1}), possibly due to the
  effective filter curve for Pristine on CFHT being slightly different
  than the assumed filter curve used to create $CaHK _ \text{synth}$.
   
  \begin{figure}
    \includegraphics{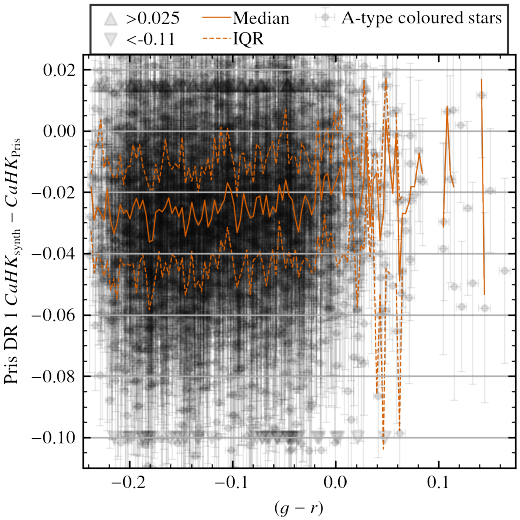}
    \caption{$CaHK$ synthetic minus $CaHK _ \text{Pris}$ for A-type
             coloured stars in the shallow dataset cut off at $-$0.110
             and 0.025 for clarity of the shown trends for median and
             interquartile range. The stars that have above 0.025 and
             below $-$0.110 are shown by triangles facing up and down
             at the top and bottom edge, respectively, of the axis at
             their respective $g\! -\! r$.}
    \label{appendix_figure_cahk_synthetic_pristine_magnitude_difference}
  \end{figure}
  
  \section{Colour-colour diagrams of candidate A-type stars}%
  \label{section_colour_colour_diagrams}
  
\begin{figure}
    \includegraphics{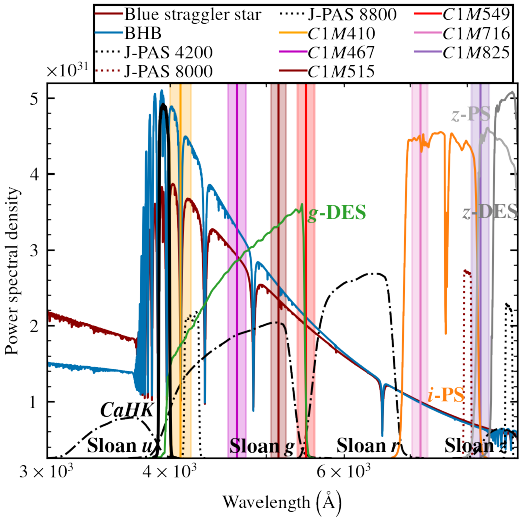}
    \caption{\citet{munari_et_al2005library} stellar library
             synthetic stellar spectra generated for stars with
             $T_{\text{eff}} = 8500 K$,
             $\text{bolometric luminosity}\ L%
             = 40\,\text{solar luminosity}$,
             $[\text{Z}/\text{H}] \simeq -0.5$. One of the stars has
             a log$_{10}(g) = 3$ value, which is typical for a BHB
             star (the blue line in the plot). The other star has a
             log$_{10}(g) = 5$ value, which is typical for a blue
             straggler (the red line in the plot). The transmission
             curves of the SDSS \textit{ugrz}
             \citep{fukugita_et_al1996sdss, Doi_2010},
             \textit{CaHK} from the Pristine survey
             \citep{Starkenburg_2017}, J-PAS 4200, 8000, and 8800
             \citep{benitez2014jpas}, DES \textit{gz}
             \citep{flaugher_et_al2015}, and Pan-STARRS (PS)
             \textit{iz} bands
             \citep{chambers_et_al_arxiv_pan_starrs1} are displayed
             and scaled to be visible. For the \textit{C1MXXX}
             filters \citep{Jordi_2006}, the centre of the band
             (represented by a solid vertical line) and the full
             width at half maximum (represented by dashed vertical
             lines) are shown.}
    \label{synthetic_stellar_spectra}
  \end{figure}

  To create alternative colour-colour spaces for BHB and blue
  straggler star discrimination, we explore various filter
  combinations using synthetic \textit{Gaia} photometry available,
  including several Javalambre Physics of the Accelerating Universe
  Astrophysical Survey (J-PAS) filters in the blue wavelengths of
  the spectrum. The aim of this exercise is to mitigate the
  relatively substantial photometric uncertainties associated with
  \textit{CaHK}. In \autoref{synthetic_stellar_spectra}, we compare
  a synthetic spectrum of a BHB star with those of a blue straggler.
  It is evident that certain features are sensitive to surface
  gravity and we choose the explored filters such that they overlap
  with these features.
    
  The J-PAS 4200, with a narrow passband peaking at 420 nm, has a
  slightly redder wavelength range compared to \textit{CaHK}.
  Despite this, it should effectively capture the flux difference
  between BHB and blue straggler stars around the Balmer jump. While
  the Pristine \textit{CaHK} filter aligns with the decrease in
  transmissivity observed in the \textit{Gaia} \textit{BP}/%
  \textit{RP} spectra, the transmissivity in the wavelength range of
  the J-PAS 4200 filter is notably higher. At similar signal-to-%
  noise, J-PAS 4200 allows us to examine the halo with BHB stars
  that are about one magnitude fainter (as shown in
  \autoref{comparison_figure}).
  
\begin{figure*}
    \includegraphics{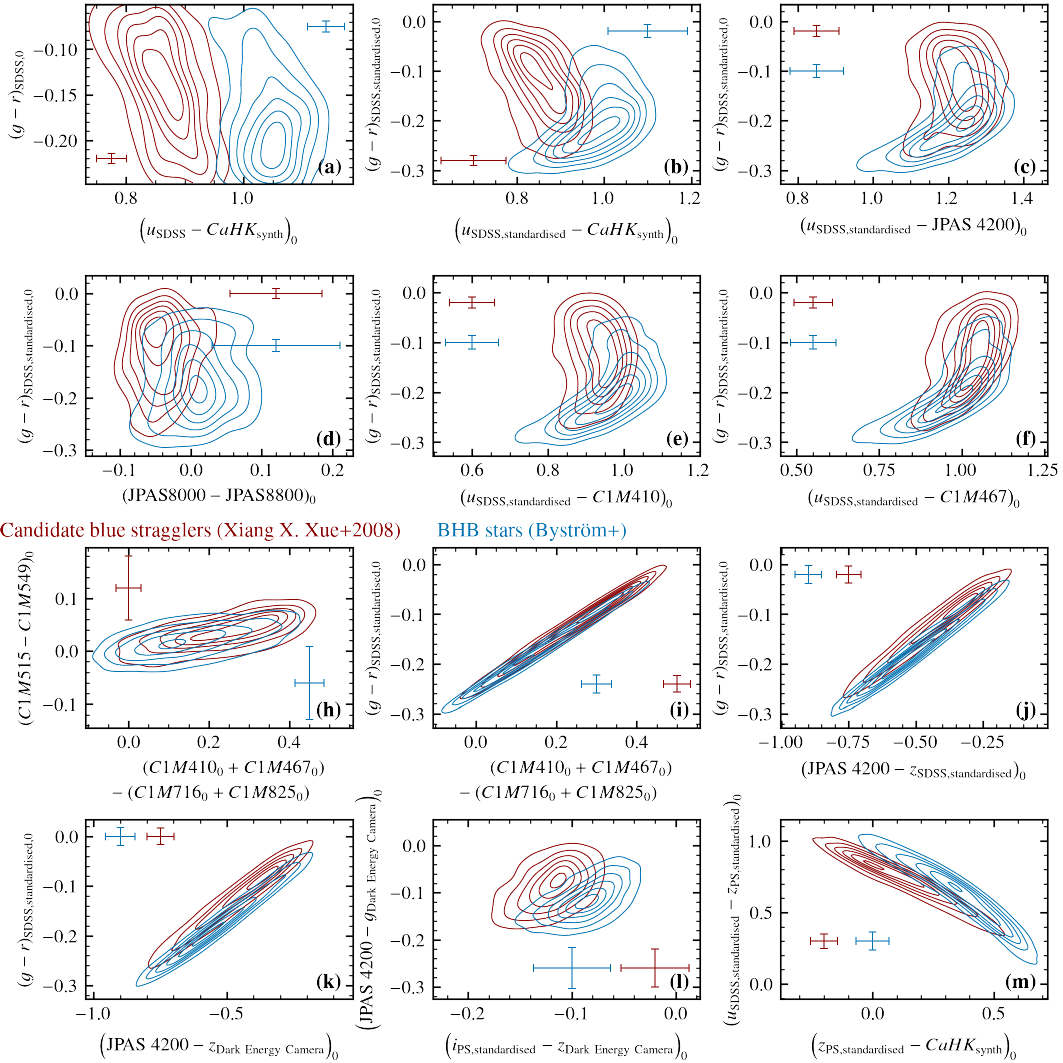}
    \caption{Comparison of the colour-colour diagram using SDSS
             photometry and synthetic \textit{CaHK} to separate BHB
             (blue) from (candidate) blue straggler stars (red) in
             the top left panel and other colour-colour diagrams
             using synthetic photometry that we explored. The
             subscript PS refers to Pan-STARRS.}
    \label{additional_colour_colour_diagrams}
  \end{figure*}

  In \autoref{additional_colour_colour_diagrams} the performance of
  $\left(u_\text{SDSS}-4200\right) _0$ and other XP synthetic
  photometry combinations are explored to roughly separate BHB from
  blue straggler stars in the full training sample. The combination
  of the J-PAS 4200 filter and \textit{ugr} photometry on the
  training sample appears to successfully discriminate between BHB
  and blue straggler stars in colour space (as shown in the top-%
  middle panel of \autoref{additional_colour_colour_diagrams}).
  However, the overlap between the two sequences in colour space is
  greater than when the Pristine \textit{CaHK} filter is used
  instead of the J-PAS 4200 filter (as shown in the top-left panel
  of \autoref{additional_colour_colour_diagrams}), lessening the
  discriminative power.
  
  We also tested redder filter combinations near the Paschen jump in
  \autoref{additional_colour_colour_diagrams}. We drew inspiration
  from works by \citet{lenz_et_al1998}, \citet{Deason2011},
  \citet{Vickers_2012}, and \citet{Starkenburg2019} (usage of
  $z _ \text{SDSS}$ in combination with J-PAS 4200 and
  $z _ \text{PS}$ and $u _ \text{SDSS}$ in combination with
  $CaHK _ \text{synth}$, as shown in panels j and m, respectively, in
  \autoref{additional_colour_colour_diagrams}), and
  \citet{Thomas2018Atype} and \citet{fengqing_yu_et_al2024} (usage
  of $\left(\text{J-PAS}\ 4200-z_\text{DES}\right) _ 0$ in
  combination with $(g-r) _ \text{SDSS,0}$ or
  $\left(i_\text{PS}-z_\text{DES}\right) _ 0$ in combination with 
  $\left(\text{J-PAS}\ 4200-g_\text{DES}\right) _ 0$ as demonstrated
  in panels k--l in \autoref{additional_colour_colour_diagrams},
  respectively). Furthermore, we tested the use of J-PAS 8000 and
  J-PAS 8800 and think that these filters might be able to stress
  the flux difference between BHB and blue straggler stars around
  the Paschen jump (see top right panel of
  \autoref{additional_colour_colour_diagrams}). From the last panel,
  we observed two branches, similar to the
  $u_\text{SDSS} - CaHK_\text{synth}$ versus
  $(g-r) _ \text{SDSS}$ case (first panel). 
  
  However, none of these panels seem to demonstrate the same
  discriminatory power as the main colour-colour space used in this
  work.
  
  \section{Corner plots for the hierarchical Bayesian mixture
           models}%
  \label{additional_corner_plots}
  
  Figure \ref{corner_plot_training_synthetic_version} summarises the
  results of the initial sampling, based on the training samples in
  shallow dataset. In addition, Figs. 
  \ref{corner_plot_synthetic_version} and \ref{corner_plot} provide
  the summary of the results of the sampling, based on the full
  samples in the shallow and deep datasets, respectively.
  
  \begin{figure*}
    \includegraphics{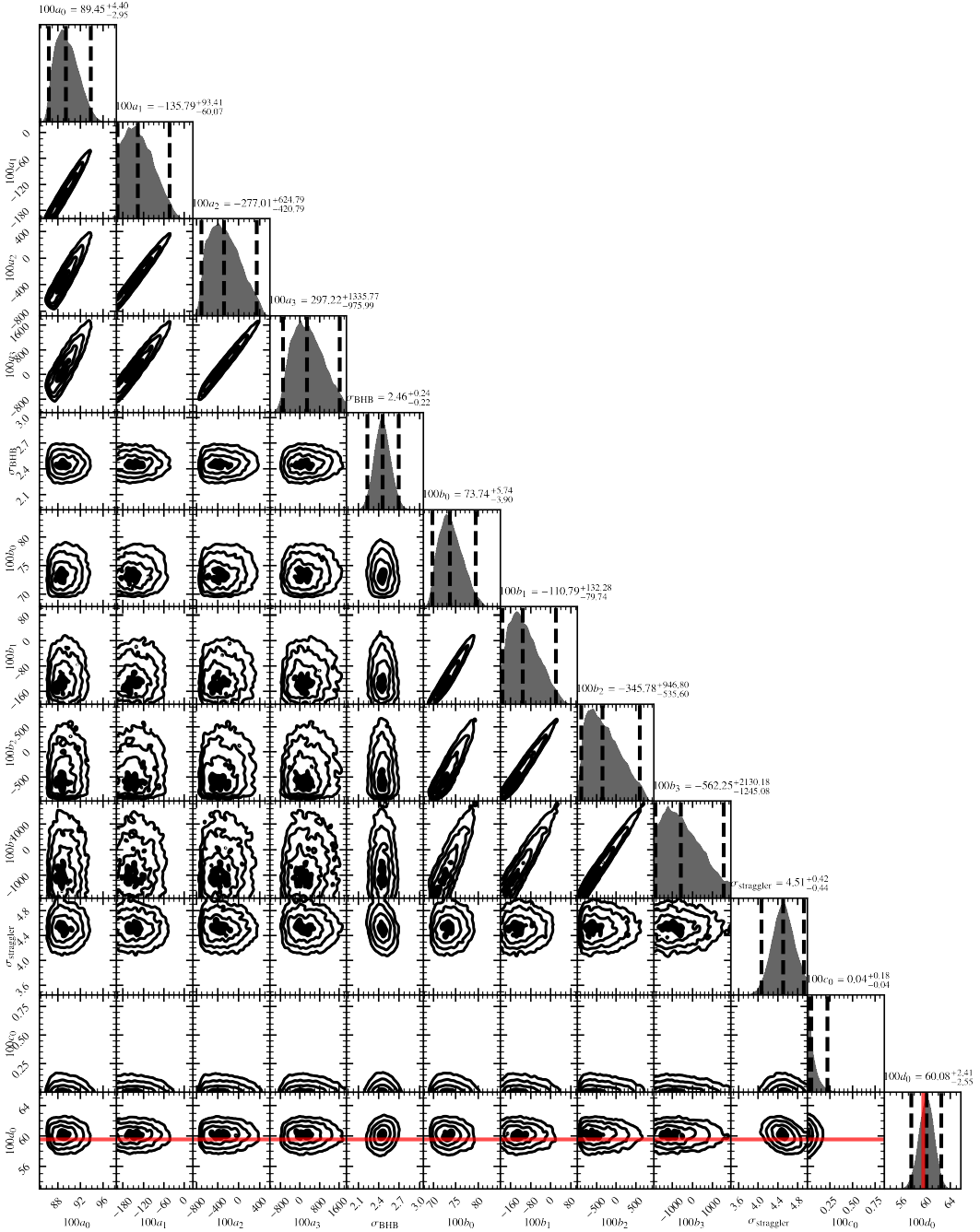}
    \caption{Two- and one-dimensional projections of the
             twelve-dimensional posterior probability distribution
             to describe the mixture of candidate BHB stars, blue
             straggler stars, and outliers for the training sample
             in the deep dataset of the
             $\left(u - CaHK_\text{Pris}\right)_0$-
             $(g-r) _ 0$ colour-colour diagram. By default, for
             numerical reasons the parameter values are scaled by
             100 in \texttt{dynesty} \citep{speagle2020} version
             2.1.3 \citep{koposov_et_al2023dynesty_version213} that
             we used here. The vertical lines and the (red)
             horizontal line in the lower panels mark the $2\sigma$
             quantiles and median of each projected one-dimensional
             distribution and the expected fraction of BHB stars in
             the training sample, respectively.}
    \label{corner_plot_training}
  \end{figure*}

  \begin{figure}
    \includegraphics{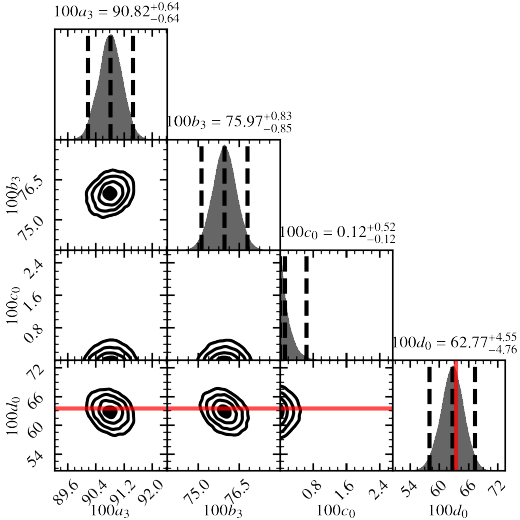}
    \caption{Two- and one-dimensional projections of the
             six-dimensional posterior probability distribution to
             describe the mixture of candidate BHB stars, blue
             straggler stars, and outliers for the training sample
             in the shallow dataset of the
             $\left(u - CaHK_\text{synth}\right)_0$-
             $(g-r) _ 0$ colour-colour diagram. By default, for
             numerical reasons the parameter values are scaled by 100
             in \texttt{dynesty} \citep{speagle2020} version 2.1.3
             \citep{koposov_et_al2023dynesty_version213} that we used
             here. The vertical lines and the (red) horizontal line in
             the lower panels mark the $2\sigma$ quantiles and median
             of each projected one-dimensional distribution and the
             expected fraction of BHB stars in the training sample,
             respectively.}
    \label{corner_plot_training_synthetic_version}
  \end{figure}
  
  \begin{figure}
    \includegraphics{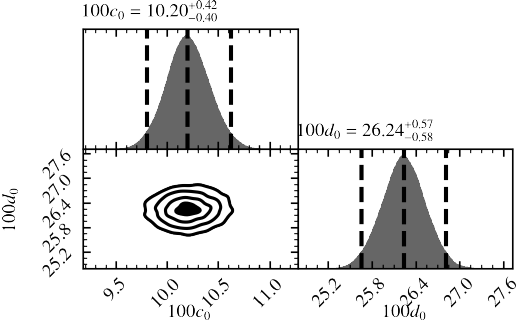}
    \caption{Equivalent version of 
             \autoref{corner_plot_training_synthetic_version} for
             the sampling of the two-dimensional posterior
             probability distribution of $c _ 0$ and $d _ 0$ given
             the full shallow dataset where we fixed $a _ 0$,
             $a _ 1$, $a _ 2$, $a _ 3$, $\sigma _ \text{BHB}$,
             $b _ 0$, $b _ 1$, $b _ 2$, $b _ 3$, and
             $\sigma _ \text{straggler}$ to the values we got from
             the sampling given the training samples.}
    \label{corner_plot_synthetic_version}
  \end{figure}
  
  \begin{figure}
    \includegraphics{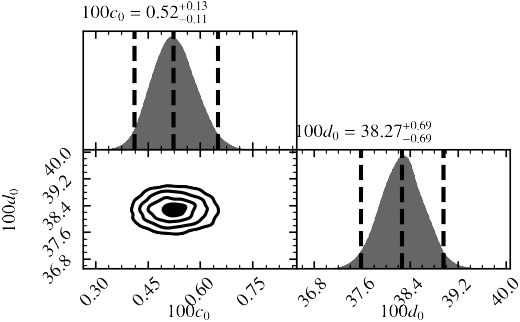}
    \caption{Equivalent version of 
             \autoref{corner_plot_training_synthetic_version} for
             the sampling of the two-dimensional posterior
             probability distribution of $c _ 0$ and $d _ 0$ given
             the full deep dataset where we fixed $a _ 0$, $a _ 1$,
             $a _ 2$, $a _ 3$, $\sigma _ \text{BHB}$, $b _ 0$,
             $b _ 1$, $b _ 2$, $b _ 3$, and
             $\sigma _ \text{straggler}$ to the values we got from
             the sampling given the training samples.}
    \label{corner_plot}
  \end{figure}
    
  \section{Average median Dark Energy Spectroscopic Instrument R arm
           signal-to-noise ratio of sample A-type candidates}%
  \label{signal_to_noise_ratio}
  
  To investigate the relative completeness estimates of the two
  samples of BHB candidates, we employed Kiel diagrams based on DESI
  MWS DR1 values from the \texttt{RVSpecFit} pipeline in
  \autoref{purity_and_completeness_figure}, following a similar
  approach as described in \citet{bystroem_et_al2025}. In
  \autoref{appendix_purity_and_completeness_figure}, we check the
  average median R arm S/N in the same Kiel
  diagrams. As anticipated, the shallow datasets that contains only
  brighter stars tend to have more precise DESI MWS DR1 R arm
  spectra on average than do stars in the deep dataset. In
  particular, stars that are classified as BHB stars following our
  method, but get higher surface gravity values in DESI, tend to
  have low S/N.
  
  \begin{figure}
    \includegraphics{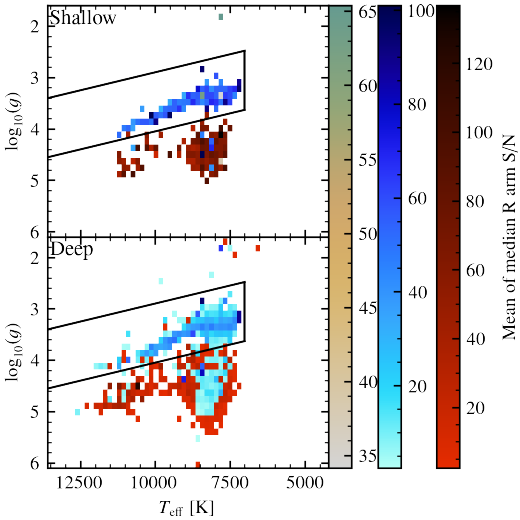}
    \caption{Two-dimensional histograms in a Kiel diagram of all
             classified good-quality stellar-like sources that have
             a effective temperature and surface gravity in DESI MWS
             DR1 from the RVSpecFit pipeline. We colour-code by
             their average median R arm S/N
             (candidate BHB stars with blue, candidate blue
             stragglers with red, and candidate outliers with
             greyish colours). To the Kiel diagrams we show the
             $1.50 + 0.00014T_\text{eff}$ and
             $2.65 + 0.00014T_\text{eff}$ lines that were used by
             \citet{bystroem_et_al2025} to select BHB stars.}
    \label{appendix_purity_and_completeness_figure}
  \end{figure}
  
  \section{Discrepant probabilities of candidate A-type stars that
           are in both datasets}%
  \label{discrepant_probabilities}
  
\begin{figure}
    \includegraphics{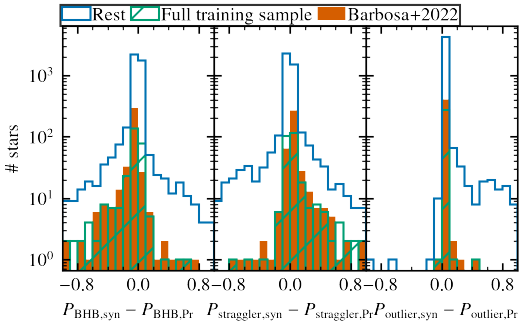}
    \caption{Histograms of $P_\text{BHB, syn} - P_\text{BHB, Pr}$,
             $P_\text{straggler, syn} - P_\text{straggler, Pr}$,
             and $P_\text{outlier, syn} - P_\text{outlier, Pr}$ from
             the hierarchical Bayesian mixture models of the 
             $\left(u-CaHK_\text{synth}\right) _ 0$ and 
             $\left(u-CaHK_\text{Pris}\right)_ 0$ vs 
             $(g-r)_0$ colour-colour diagrams of candidate A-type stars
             for stars that are both in the shallow and deep
             datasets. We split the two datasets into the stars that
             are in the full training sample, extra BHB stars with
             100\% probability of being a BHB star in the
             \citet{barbosa_et_al2022} sample, and the rest of the two
             datasets.}
    \label{histograms_probability_comparison}
  \end{figure}

  Figure \ref{histograms_probability_comparison} presents a
  comparison for stars present in both the shallow and deep samples
  of being a BHB, blue straggler, or outlier candidate between both
  models. The histograms peak very steeply around zero difference of
  the population probabilities between the synthetic and Pristine
  datasets. However, the $P_\text{BHB, syn} - P_\text{BHB, Pr}$ and
  $P_\text{straggler, syn} - P_\text{straggler, Pr}$ 
  histograms show wings on both sides of 0.0 that reach -1.0 and 1.0
  while the $P_\text{outlier, syn} - P_\text{outlier, Pr}$ histogram
  has only one wing towards 1.0.
  
  We have investigated whether objects
  that are outlying have proportionally higher uncertainties in
  $\delta (g-r)_0$ and/or
  $\delta\left(u-CaHK_\text{synth}\right) _ 0$
  (with the note that $\left(u-CaHK_\text{synth}\right) _ 0$ will be on
  average more uncertain than $\left(u-CaHK_\text{Pris}\right) _ 0$).
  Moreover, we also checked whether there is any $(g-r) _ 0$ dependence
  on varying classifications. Neither of these trends was found: we do
  not see a significant increase in $\delta (g-r)_0$ and/or
  $\delta\left(u-CaHK_\text{synth}\right) _ 0$ for objects with
  $P_\text{BHB, syn} - P_\text{BHB, Pr}$,
  $P_\text{straggler, syn} - P_\text{straggler, Pr}$, and
  $P_\text{outlier, syn} - P_\text{outlier, Pr}$ off zero, nor do we see
  an apparent correlation between the probability differences and
  $(g-r) _ 0$. A thorough investigation is beyond the scope of this
  paper, but we suspect that the remaining outliers are caused by a
  combination of factors. For some objects, it could be related to some
  leftover variability. We furthermore presume the one-sided wing
  towards 1.0 in the $P_\text{outlier, syn} - P_\text{outlier, Pr}$
  histogram could be due to a larger fraction of outlier candidates in
  the modelling of that dataset.
  
  \section{Description of the data table}\label{tableDescription}
  
  The final table, available online or upon request, includes
  \textit{Gaia} identification data (\textit{Gaia} DR3 ID and
  coordinates), de-reddened photometric colours, maximum likelihood
  probabilities for BHB, blue straggler, and outlier classes, and
  inferred distances (see \autoref{dataset_description}). Line-of-sight
  velocities, Galactocentric positions, angular momentum, and energies
  in two Galactic gravitational potentials are also included for
  scientific explorations. A flag indicates whether the BHB star could
  be identified with a known substructure.
  
  \begin{table*}
    \caption{Description of the columns of the combined shallow and
             deep datasets of candidate A-type stars catalogue that are
             available at the CDS.}
    \label{dataset_description}
    \begin{tabular}{ll}
      \hline\hline
      Column & Description\\
      \hline
      source\_id & \textit{Gaia} DR3 source ID\\
      RA & Right ascension\\
      Dec & Decl.\\
      source\_catalogue & Source catalogue\\
      sfd\_ebv & $E\left(B-V\right)$\\
      u0\_sdss\_cahk0\_synth & $(u - CaHK_\text{synth})_0$\\
      du0\_sdss\_cahk0\_synth
      & $\delta(u - CaHK_\text{synth})_0$\\
      u0\_sdss\_cahk0\_pris & $(u - CaHK_\text{Pris})_0$\\
      du0\_sdss\_cahk0\_pris & $\delta(u - CaHK_\text{Pris})_0$%
      \\
      gr0\_sdss & $(g-r) _ 0$\\
      dgr0\_sdss & $\delta(g-r) _ 0$\\
      maximum\_likelihood\_probability\_bhb\_star\_syn
      & $P _ \text{BHB,synth}$\\
      maximum\_likelihood\_probability\_straggler\_syn
      & $P _ \text{straggler,synth}$\\
      maximum\_likelihood\_probability\_outlier\_syn
      & $P _ \text{outlier,synth}$\\
      maximum\_likelihood\_probability\_bhb\_star\_pr
      & $P _ \text{BHB,Pris}$\\
      maximum\_likelihood\_probability\_straggler\_pr
      & $P _ \text{straggler,Pris}$\\
      maximum\_likelihood\_probability\_outlier\_pr
      & $P _ \text{outlier,Pris}$\\
      distance & $D _ \text{h}$[kpc]\\
      distance\_uncertainty & Uncertainty of $D _ \text{h}$[kpc]\\
      rv\_best & $v _ \text{h}\left[\text{km}\,\text{s}^{-1}\right]$\\
      rv\_best\_err
      & $\delta v _ \text{h}\left[\text{km}\,\text{s}^{-1}\right]$\\
      rv\_source  & Source catalogue for line-of-sight velocity\\
      rv\_source\_id & ID in line-of-sight velocity source catalogue\\
      x & Galactocentric X[kpc]\\
      x\_uncertainty & Uncertainty of Galactocentric X[kpc]\\
      y & Galactocentric Y[kpc]\\
      y\_uncertainty & Uncertainty of Galactocentric Y[kpc]\\
      z & Galactocentric Z[kpc]\\
      z\_uncertainty & Uncertainty of Galactocentric Z[kpc]\\
      energy1
      & $E\left[\text{km}^2\,\text{s}^{-2}\right]$ in setup
      (1)\\
      denergy1
      & Uncertainty of 
      $E\left[\text{km}^2\,\text{s}^{-2}\right]$ in setup (1)%
      \\
      LZ1 &
      $L _ Z\left[\text{kpc}\,\text{km}\,\text{s}^{-1}\right]$
      in setup (1)\\
      dLZ1
      & Uncertainty of
      $L _ Z\left[\text{kpc}\,\text{km}\,\text{s}^{-1}\right]$
      in setup (1)\\
      energy2
      & $E\left[\text{km}^2\,\text{s}^{-2}\right]$ in setup 
      (2)\\
      denergy2
      & Uncertainty of 
      $E\left[\text{km}^2\,\text{s}^{-2}\right]$ in setup (2)%
      \\
      LZ2
      &
      $L _ Z\left[\text{kpc}\,\text{km}\,\text{s}^{-1}\right]$
      in setup (2)\\
      dLZ2
      & Uncertainty of
      $L _ Z%
      \left[\text{kpc}\,\text{km}\,\text{s}^{-1}\right]$
      in setup (2)\\
      Lperpendicular
      &
      $L _ \perp%
      \left[\text{kpc}\,\text{km}\,\text{s}^{-1}\right]$
      in setup (2)\\
      dLperpendicular
      & Uncertainty of
      $L _ \perp%
      \left[\text{kpc}\,\text{km}\,\text{s}^{-1}\right]$
      in setup (2)\\
      LperpendicularZ2\_covariance
      & $L _ Z$-$L _ \perp$ covariance%
      $\left[\text{kpc}^2\,\text{km}^2\,\text{s}^{-2}\right]$ in setup 
      (2)\\
      energyLZ2\_covariance
      & $E$-$L _ Z$ covariance%
      $\left[\text{kpc}\,\text{km}^3\,\text{s}^{-3}\right]$ in setup
      (2)\\
      energy2Lperpendicular\_covariance
      & $E$-$L _ \perp$ covariance%
      $\left[\text{kpc}\,\text{km}^3\,\text{s}^{-3}\right]$ in setup
      (2)\\
      group & Group from substructure assignment\\
      \hline\hline
    \end{tabular}
    \tablefoot{The equatorial coordinates are from \textit{Gaia} DR3.
               The possible values of source\_catalogue are:
               Pristine-\textit{Gaia} synthetic catalogue
               (pristine\_gaia\_sdss value), Pristine survey
               (pristine\_sdss value), or both (both value).
               $E\left(B-V\right)$ is from
               \citet{Schlegel+98}. The heliocentric distance and its
               uncertainty are NaN if $P_\text{BHB} \leq 0.5$. The
               possible values of rv\_source are: SEGUE,
               LAMOST\_MRS\_gen, LAMOST\_LRS\_stellar, GAIA,
               LAMOST\_MRS\_stellar, LAMOST\_LRS\_gen, GALAH, and
               DESI. Two setups are described: one with Cartesian
               coordinates centred on the Galactic centre for
               $Z_\odot = 0.014$\,kpc and the \citet{mcmillan2017}
               gravitational potential, and another with coordinates
               centred on the Galactic centre for
               $Z_\odot = 0.000$\,kpc and the gravitational potential
               used in \citet{Dodd_2023}. The possible values
               of group are: 0 (Gaia-Enceladus), 1
               \cite[L-RL3;][]{L_vdal_2022}, 2 (hot thick disc), 3
               \cite[Thamnos 1 and 2;][]{koppelman_et_al2019_631_l9}, 4
               \cite[Helmi streams;][]{helmi_et_al1999helmi_streams}, 5
               \cite[Sequoia;][]{myeong_et_al2019}, 6 (ED-1), 13, 14,
               16, 17 (ED-4), 20, 21, 23 (ED-3), 28 (ED-6), other, and
               sagittarius. The numbers refer to the clusters
               (re)identified by \citet{Dodd_2023}.}
  \end{table*}
  
  \section{Removal of four uncertain stars}\label{star_removal}
  From our samples we removed two BHB candidates in the shallow dataset
  that turned out later in the analysis of the angular momenta and
  energy to be misclassification due to their unrealistically great
  angular momenta and energies (positive energies) which were caused by
  placing the stars at the wrong distance due to their
  misclassification. In support of their misclassifications, BHB
  candidate \textit{Gaia} (E)DR3 834359077274173184 is given only low
  probabilities of being a BHB candidate in
  \citet[][0.12661]{k_w_smith_et_al2010} and
  \citet[][$1.5000\times 10^{-7}$]{amarante_et_al2024}, and SDSS object
  1237665231601008859 has $T_\text{eff} = 8116\!\pm\! 6$\,K and
  $\log_{10} = 4.21\!\pm\!0.02$ in the DESI MWS DR1 and a
  probability of 0.05119 of being a BHB candidate in
  \citet{amarante_et_al2024}. 
  
  We also did not consider BHB candidates that have a missing
  line-of-sight velocity uncertainty, Gaia (E)DR3 2573593817362548736
  because the available redshift estimate in the LAMOST low-resolution
  spectroscopic general DR11 version 2.0 catalogue has an uncertainty
  of 50 and is also classified as a white dwarf star in the same
  catalogue, and SDSS object 1237662663218495613 since its
  line-of-sight velocity estimate is based on an unreliable blue LAMOST
  mid-resolution DR11 version 2.0 spectrum.
    
  \section{Comparison of shallow dataset to other samples of blue
           horizontal branch candidates selected based on synthetic
           photometry}
  \label{other_catalogues}
         
  To assess the significance of our catalogue of BHB candidates in the
  shallow dataset compared to other recent studies on selecting BHB
  candidates using synthetic photometry, we conducted a comparison with
  two recent catalogues of BHB candidates mentioned in \autoref{intro},
  namely \citet{guozhen_hu_et_al2025} and \citet{wenbo_wu_et_al2025}.
  \citet{guozhen_hu_et_al2025} built upon the SkyMapper Southern Sky
  Survey standard catalogue, which is included in the BEst STar (BEST)
  database \citep{kai_xiao_et_al2026}. Their initial sample comprised
  29,232,880 stars that met specific criteria. These criteria included
  having a flag value of 000/001/002 for the corrected Gaia
  spectrophotometry \citep{bowen_huang_et_al2024}, photometric
  uncertainties below 0.04 magnitude, and Galactic latitudes of eight
  degrees or higher. To select the candidate BHB stars, they employed
  Gaussian mixture modelling of the $(u-v) _ {\text{SkyMapper},0}$
  histogram across various bins of $(g-i) _ {\text{SkyMapper},0}$ and
  Galactic latitudes for $0.8\leq (u-v)_{\text{SkyMapper},0}\leq 1.9$
  and $-0.50\leq (g-i)_{\text{SkyMapper},0}\leq -0.08$ (figures 1 and
  B1—B6).
  
  The BEST database is based on corrected \textit{Gaia}
  spectrophotometry that takes into account the influence of the
  Earth's atmosphere, optical system, filters, and detectors
  \citep{bowen_huang_et_al2024}. For near-ultraviolet filter(s)
  an extrapolation based on a linear function is required. The
  corrections to the \textit{Gaia} spectrophotometry were estimated from
  a comparison of a set of reference, stellar spectra from CALSPEC
  \citep{bohlin2014}, the next generation spectral library
  \citep{heap_and_lindler2007}, the medium resolution Isaac Newton
  telescope library of empirical spectra
  \citep{sanchez_blazquez_et_al2006}, and LEMONY, and \textit{Gaia} DR3
  spectrophotometry in the default wavelength range 336--1020\,nm in
  steps of 2\,nm in the \textit{Gaia} DR3 spectrophotometry; that is,
  the reference, stellar spectra are processed (convolution, and inter-
  and extrapolation) to match the \textit{Gaia} DR3 spectrophotometry.
  
  \citet{wenbo_wu_et_al2025} employed FERRE
  \citep{allende_prieto_et_al2006} instead to model the XP spectra of
  stars having synthetic standardised
  $0.8 < (u-g) _ \text{SDSS,0} < 1.4$ and
  $-0.3 < (g-r) _ \text{SDSS,0} < 0.1$,
  $\left(u_\text{SDSS}-CaHK_\text{synth}\right) _ 0$
  consistent with being a BHB star (figure 1), $G > 11.5$ to cut stars
  that may have some XP spectra instrumental effects from the internal
  calibration process as mentioned by \citet{andrae_et_al2023}, and
  Galactic longitudes and latitudes outside of
  $274.5\degr < l < 286.5\degr$, $-37.9\degr < b < -27.9\degr$,
  $299.8\degr < l < 305.8\degr$, and $-46.3\degr < b < -42.3\degr$ to
  remove stars in the regions of the Large and Small Magellanic Clouds,
  respectively, with degraded synthetic spectra from the nsc3 library
  \citep{allende_prieto_et_al2018} to a constant spectral resolving
  power of 104. They employed additionally a 0.1 cut on the fitting
  residuals, $|b| > 20\degr$, and a final Kiel diagram selection
  (figure 2) to end with 44,552 BHB candidates.
  
\begin{figure}
    \includegraphics{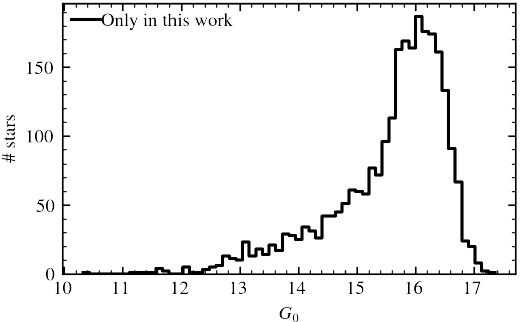}
    \caption{$G _ 0$ of all exclusive BHB candidates from our work.}
    \label{histogram_dereddened_gaia_magnitude}
  \end{figure}

  We show the $G _ 0$ of each exclusive BHB candidate in our shallow
  dataset in a histogram in
  \autoref{histogram_dereddened_gaia_magnitude}. We used the subset of
  the shallow dataset as described before the science results,
  meaning we do not have the $E\left(B-V\right) < 0.1$ cut here.
  $G _ 0$ was estimated using the \citet{Schlegel+98}
  $E\left(B-V\right)$ values in each catalogue and
  $\frac{A _ G}{A _ {V}} = 0.83627$
  \citep{cardelli_et_al1989}.
         
  If we take the catalogues by \citet{guozhen_hu_et_al2025} and
  \citet{wenbo_wu_et_al2025} at face value without any quality cuts,
  we find that as anticipated we add BHB candidates to the other
  catalogues towards the available faint part of the set of
  \textit{Gaia} DR3 sources for which XP spectra are available (compare
  to figure 5 in \citealt{wenbo_wu_et_al2025}) because we used the deep
  SDSS \textit{u} data. With the added caveat that the quality of the
  photometry and possibly probabilistic photometric classification will
  decrease with increasing/decreasing reddening/$b$, we would like to
  note in terms of sky coverage that we add BHB candidates for
  $|b| < 8\degr$. Therefore, excluding this part, the shallow dataset
  of BHB candidates really helps in adding faint BHB candidates in
  comparison to the works by \citet{guozhen_hu_et_al2025} and
  \citet{wenbo_wu_et_al2025}.

\end{appendix}

\end{document}